\pdfoutput=1
\RequirePackage{ifpdf}
\ifpdf 
\documentclass[pdftex]{sigma}
\else
\documentclass{sigma}
\fi

\usepackage{stmaryrd}
\numberwithin{equation}{section}

\begin{document}

\allowdisplaybreaks

\renewcommand{\thefootnote}{$\star$}

\renewcommand{\PaperNumber}{051}

\FirstPageHeading

\ShortArticleName{Gravity in Twistor Space and~its Grassmannian Formulation}

\ArticleName{Gravity in Twistor Space\\
and~its Grassmannian Formulation\footnote{This paper is a~contribution to the Special Issue on Progress in Twistor Theory.
The full collection is available at \href{http://www.emis.de/journals/SIGMA/twistors.html}{http://www.emis.de/journals/SIGMA/twistors.html}}}

\Author{Freddy CACHAZO~$^{\dag}$, Lionel MASON~$^{\ddag}$ and David SKINNER~$^{\S}$}

\AuthorNameForHeading{F.~Cachazo, L.~Mason and D.~Skinner}

\Address{${}^{\dag}$~Perimeter Institute for Theoretical Physics,\\
\hphantom{${}^{\dag}$}~31~Caroline~St., Waterloo, Ontario N2L 2Y5, Canada}
\EmailD{\href{mailto:fcachazo@perimeterinstitute.ca}{fcachazo@perimeterinstitute.ca}} 

\Address{${}^{\ddag}$~The Mathematical Institute, 24-29~St.~Giles', Oxford OX1 3LB, UK}
\EmailD{\href{mailto:lmason@maths.ox.ac.uk}{lmason@maths.ox.ac.uk}}

\Address{${}^{\S}$~DAMTP, Centre for Mathematical Sciences, Wilberforce Road, Cambridge CB3 0WA, UK}
\EmailD{\href{mailto:d.b.skinner@damtp.cam.ac.uk}{d.b.skinner@damtp.cam.ac.uk}}

\ArticleDates{Received November 21, 2013, in f\/inal form April 23, 2014; Published online May 01, 2014}

\Abstract{We prove the formula for the complete tree-level $S$-matrix of~$\mathcal{N}=8$ supergravity recently conjectured by two of the authors.
The proof proceeds by showing that the new formula satisf\/ies the same BCFW recursion relations that physical amplitudes are known to satisfy,
with the same initial conditions.
As part of the proof, the behavior of the new formula under large BCFW deformations is studied.
An unexpected bonus of the analysis is a~very straightforward proof of the enigmatic $1/z^2$ behavior of gravity.
In addition, we provide a~description of gravity amplitudes as a~multidimensional contour integral over a~Grassmannian.
The Grassmannian formulation has a~very simple structure; in the N$^{k-2}$MHV sector the integrand is essentially the product of that of an MHV
and~an $\overline{{\rm MHV}}$ amplitude, with $k+1$ and~$n-k-1$ particles respectively.}


\Classification{53C28}

\renewcommand{\thefootnote}{\arabic{footnote}}
\setcounter{footnote}{0}

\section{Introduction}

In a~recent paper~\cite{Cachazo:2012kg}, two of us conjectured that the complete classical $S$-matrix of maximal supergravity in four dimensions
can be described by a~certain integral over the space of rational maps to twistor space.
The main aim of this paper is to prove that conjecture.

In~\cite{ArkaniHamed:2008gz,ArkaniHamed:2008yf, Bedford:2005yy, Benincasa:2007qj,Cachazo:2005ca,Cheung:2008dn} it was shown that gravitational
tree amplitudes obey the BCFW recursion relations~\cite{Britto:2004ap,Britto:2005fq}.
Our method here is to show that the formula presented in~\cite{Cachazo:2012kg} obeys these same relations, and~produces the correct
three-particle MHV and~$\overline{\mathrm{MHV}}$ amplitudes to start the recursion.

In the analogous formulation of tree amplitudes in $\mathcal{N}=4$ super Yang--Mills~\cite{Berkovits:2004hg,Roiban:2004yf, Witten:2003nn}, BCFW
decomposition is closely related to performing a~contour integral in the moduli space of holomorphic maps so as to localize on the boundary
where the worldsheet degenerates to a~nodal
curve~\cite{Bourjaily:2010kw,Dolan:2009wf,Dolan:2011za,Gukov:2004ei,Skinner:2010cz,Spradlin:2009qr, Vergu:2006np}.
The various summands on the right hand side of the recursion relation correspond to the various ways the vertex operators and~map degree may be
distributed among the two curve components.

The relation between factorization channels of amplitudes and~shrinking cycles on the worldsheet that separate some vertex operators from others
is of course a~fairly general property of string theory.
In the present case, it is also necessary to prove that the rest of the integrand behaves well under this degeneration.
In particular, the formula of~\cite{Cachazo:2012kg} involves a~product of two determinants that generalize Hodges' construction of gravitational
MHV amplitudes~\cite{Hodges:2011wm,Hodges:2012ym} to arbitrary external helicities.
One of the striking properties of these determinants is that they each depend in a~simple way on the `inf\/inity twistor', and~thereby the
breaking of conformal invariance inherent in gravitational amplitudes becomes completely explicit.
More specif\/ically, the determinants are each monomials in the inf\/inity twistor, to a~power that depends only on the number of external states
and~the MHV degree of the scattering process.
Furthermore, as explained in~\cite{Cachazo:2012kg}, the way arbitrary gravitational amplitudes depend on the inf\/inity twistor can easily be
traced through BCFW recursion.
This strongly suggests that the determinants behave simply under factorization.
We shall see that this is indeed the case.

Along the way, we show that the $1/z^2$ decay~\cite{ArkaniHamed:2008yf,Cheung:2008dn} of gravitational tree amplitudes at large values of the
BCFW shift parameter~$z$ is also simple to see from the formula of~\cite{Cachazo:2012kg}.
This behaviour is responsible for many remarkable properties of these amplitudes (see, e.g.,~\cite{Spradlin:2008bu} for some applications).

In addition, in the second part of the paper we reformulate the construction of~\cite{Cachazo:2012kg} as an integral over the Grassmannian
G$(k,n)$, written in terms of the `link' coordinates of~\cite{ArkaniHamed:2009si}.
As preparation, we show how the two determinants, which in twistor space look very dif\/ferent from one another, are naturally conjugate under
parity.
The formulation of gravitational tree amplitudes as an integral over G$(k,n)$ is strikingly simple: the integrand is the product of that of an
MHV and~an $\overline{{\rm MHV}}$ amplitude, with $k+1$ and~$n-k-1$ particles respectively.

\section{Gravity from rational curves}

We begin by brief\/ly reviewing the conjecture of~\cite{Cachazo:2012kg} (see Appendix~\ref{sec:conventions} for notation).
All~$n$-particle tree amplitudes in $\mathcal{N}=8$ supergravity are given by the sum
\begin{gather*}
\mathcal{M}_n=\sum\limits_{d=0}^\infty \mathcal{M}_{n,d}
\end{gather*}
over N$^{d-1}$MHV partial amplitudes.
The main claim of~\cite{Cachazo:2012kg} is that these N$^{d-1}$MHV amplitudes may be represented by the integral
\begin{gather}
\mathcal{M}_{n,d}\big(\big\{\lambda_i,\tilde\lambda_i,\eta_i\big\}\big)
= \int \frac{\prod\limits_{a=0}^d \mathrm{d}^{4|8}\mathcal{Z}_a}{{\rm vol\,GL(2;\mathbb{C })}}|\Phi|'
|\widetilde{\Phi}|' \prod\limits_{i=1}^n (\sigma_i\mathrm{d}\sigma_i)h_i(\mathcal{Z}(\sigma_i)).
\label{gravOld}
\end{gather}
Here, $\mathcal{Z}$ is a~holomorphic map from a~rational curve $\Sigma$ with homogeneous coordinates $\sigma^{\underline\alpha}$ to
$\mathcal{N}=8$ supertwistor space with homogeneous coordinates $\mathcal{Z}^I=(Z^{\rm a},\chi^A)=(\mu^{\dot\alpha},\lambda_\alpha,\chi^A)$.
The external states are $\mathcal{N}=8$ linearized supergravitons, and~are described on twistor space by classes $h_i\in
H^{0,1}(\mathbb{PT},\mathcal{O}(2))$.
These wavefunctions are pulled back to points $\sigma_i$ on the curve via the map $\mathcal{Z}$.
We usually take
\begin{gather}
h_i(\mathcal{Z}(\sigma_i)):= \int\frac{\mathrm{d} t_i}{t_i^3}\delta^2(\lambda_i-t_i\lambda(\sigma_i))\exp(t_i\llbracket\mu(\sigma_i)\tilde\lambda_i\rrbracket)
\label{momeig}
\end{gather}
so as to represent momentum eigenstates\footnote{In~\cite{Cachazo:2012kg} we used a~G$(2,n)$ notation for the worldsheet variables, whereas here
we are working projectively on the worldsheet.
The two pictures are related by taking $\sigma_i^{\rm there} = t_i^{1/d} \sigma_i^{\rm here}$ when $d\geq1$, so that $\mathrm{d}^2\sigma_i^{\rm
there}= \frac{1}{d}(\sigma_i\mathrm{d}\sigma_i)^{\rm here}\mathrm{d} t_i/t_i$.
When $d=0$ the relation is simply $\sigma_i^{\rm there} =\sigma_i^{\rm here}$ and~$\mathrm{d}^2\sigma_i^{\rm
there}=(\sigma_i\mathrm{d}\sigma_i)^{\rm here}\mathrm{d} t_i/t_i$.
This rescaling also accounts for the factors of~$t_i$ that appear in the matrices~$\Phi$ and~$\widetilde{\Phi}$ in~\eqref{phidef}
and~\eqref{tphidef}, respectively.}.
It is easy to check that such an $h_i(\mathcal{Z}(\sigma_i))$ is homogeneous of degree $-4$ in the external data $\lambda_i$, as required for
a~positive helicity graviton supermultiplet in on-shell momentum space.

The main content of~\eqref{gravOld} is the operators $|\Phi|'$ and~$|\tilde\Phi|'$.
These arise as a~generalization of Hodges' formulation of MHV amplitudes~\cite{Hodges:2011wm,Hodges:2012ym} and~were def\/ined
in~\cite{Cachazo:2012kg} as follows.
We f\/irst let $\widetilde{\Phi}$ be the $n\times n$ matrix operator with elements
\begin{gather}
\widetilde{\Phi}_{ij} = \frac{1}{(ij)}\left[\frac{\partial}{\partial\mu(\sigma_i)}\frac{\partial}{\partial\mu(\sigma_j)}\right] =
\frac{[ij]}{(ij)}t_it_j
\qquad
\text{for}
\quad
i\neq j,
\nonumber
\\
\widetilde{\Phi}_{ii} = -\sum\limits_{j\neq i} \widetilde{\Phi}_{ij} \prod\limits_{a=0}^{d}\frac{(jp_a)}{(ip_a)},
\label{tphidef}
\end{gather}
where the second equality in the f\/irst line follows when $\widetilde{\Phi}$ acts on the momentum eigenstates~\eqref{momeig}.
It was shown in~\cite{Cachazo:2012kg} that $\widetilde{\Phi}$ has rank $n-d-2$, with the $(d+2)$-dimensional kernel spanned by vectors whose
$j^{\rm th}$ component is $\sigma_j^{\underline\alpha_0}\cdots\sigma_j^{\underline\alpha_d}$, i.e.\
\begin{gather*}
\sum\limits_{j=1}^n\widetilde{\Phi}_{ij}\sigma_j^{\underline\alpha_0}\cdots\sigma_j^{\underline\alpha_d} = 0.
\end{gather*}
This equation holds on the support of the $\delta$-functions $\prod\limits_{a=0}^d\delta^{2|8}\left(\sum\limits_i
t_i\tilde\lambda_i\big(\sigma_i^{\underline 1}\big)^a \big(\sigma_i^{\underline 2}\big)^{d-a}\right)$ obtained by integrating out the map coef\/f\/icients $\mu_a$
from~\eqref{N=8}.

$\Phi$ is similarly def\/ined as the symmetric $n\times n$ matrix with elements
\begin{gather}
\Phi_{ij} = \frac{\langle \lambda(\sigma_i)\lambda(\sigma_j)\rangle}{(ij)} = \frac{\langle ij\rangle}{(ij)}\frac{1}{t_it_j}
\qquad
\text{for}
\quad
i\neq j,
\nonumber
\\
\Phi_{ii} = -\sum\limits_{j\neq i} \Phi_{ij} \prod\limits_{r=0}^{\tilde d}\frac{(jp_r)}{(ip_r)}\frac{\prod\limits_{k\neq
i}(ik)}{\prod\limits_{l\neq j}(jl)},
\label{phidef}
\end{gather}
where
\begin{gather*}
\tilde d:= n-d-2
\end{gather*}
is introduced for later convenience, and~again the second equality in the f\/irst line of~\eqref{phidef} follows when acting on~\eqref{momeig}.
$\Phi$ has rank~$d$, with kernel def\/ined by the equation
\begin{gather}
\sum\limits_{j=1}^n \Phi_{ij} \frac{\sigma_j^{{\underline\alpha}_0} \cdots\sigma_j^{{\underline\alpha}_{\tilde d}}}{\prod\limits_{k\neq j} (jk)}
= 0.
\label{kerphi}
\end{gather}
This equation holds because for any degree~$d$ polynomial $\lambda(\sigma)$, the residue theorem gives
\begin{gather*}
\oint_{|\langle i\sigma\rangle|=\epsilon} (\sigma\mathrm{d}\sigma)
\frac{\lambda(\sigma)\sigma^{\underline\alpha_1}\cdots\sigma^{\underline\alpha_{\tilde d}}}{\prod\limits_{j=1}^n(\sigma j)} =
\frac{\lambda(\sigma_i) \sigma_i^{\underline\alpha_1}\cdots\sigma_i^{\underline\alpha_{\tilde d}}}{\prod\limits_{j\neq i}(ij)}.
\end{gather*}
The sum of this residue over all $i\in\{1,\ldots,n\}$ vanishes because the resulting contour is homologically trivial.

The operators $|\Phi|'$ and~$|\widetilde{\Phi} |'$ are def\/ined as follows.
Remove any $d+2$ rows and~any $d+2$ columns from $\tilde\Phi$ to produce a~non-singular matrix $\tilde\Phi_{\rm red}$.
Then
\begin{gather*}
|\widetilde{\Phi}|':= \frac{\det (\tilde\Phi_{\rm red})}{|\tilde r_1\dots {\tilde r}_{d+2}||\tilde c_1 \dots \tilde c_{d+2}|}.
\end{gather*}
In this ratio, $|\tilde r_1\dots \tilde r_{d+2}|$ denotes the Vandermonde determinant
\begin{gather*}
|\tilde r_1\dots \tilde r_{d+2}| = \prod\limits_{\substack{i<j
\\
i,j\in \{{\rm removed}\}}}(i j)
\end{gather*}
made from all possible combinations of the worldsheet coordinates corresponding to the deleted rows; $|\tilde c_1 \dots \tilde c_{d+2}|$ is the
same Vandermonde determinant, but for the deleted columns.

In~\cite{Cachazo:2012kg}, $|\Phi|'$ was also def\/ined in terms of the determinant of a~non-singular matrix $\Phi_{\rm red}$ obtained by similarly
removing rows and~columns from~$\Phi$, now $n-d$ of each.
However, $|\Phi|'$ itself was constructed as
\begin{gather*}
|\Phi|':= \frac{\det (\Phi_{\rm red})}{|r_1\dots r_d||c_1\dots c_d|}
\end{gather*}
using the Vandermonde determinants $|r_1\dots r_d|$ and~$|c_1\dots c_d|$ of the rows and~columns that {\it remain} in $\Phi_{\rm red}$.

\subsection[Definition of $\det'$]{Definition of $\boldsymbol{\det'}$}

The above def\/initions of~$|\widetilde{\Phi}|'$ and~$|\Phi|'$ are actually quite dif\/ferent.
This motivates us to f\/ind a~more canonical way to def\/ine these determinants.
It turns out that the most natural def\/inition has to do with the null vectors of each matrix.
Once the null space of any symmetric $n\times n$ matrix~$K$ of rank~$m$ is determined, one can compute any two maximal minors of the $n\times
(n-m)$ matrix with the null vectors of~$K$ as its columns.
Denote the two maximal minors chosen by~$|R|$ and~$|C|$.
Then
\begin{gather*}
{\det}'(K) = \frac{\det (K_{\rm red})}{|R||C|},
\end{gather*}
where $K_{\rm red}$ is the reduced matrix obtained after removing $n-m$ rows and~columns whose row and~column label coincide with the ones
removed from the $n\times (n-m)$ matrix of null vectors to obtain $|R|$ and~$|C|$.

Appendix A contains a~formal motivation for this def\/inition and~explains how the old and~new def\/initions are related.
At this point it suf\/f\/ices to say that
\begin{gather*}
|\tilde\Phi|' = {\det}'(\widetilde{\Phi})
\qquad
\text{while}
\quad
|\Phi|' = \frac{\det'(\Phi)}{|1\dots n|^2},
\end{gather*}
so that an alternative presentation of the gravity formula~\eqref{gravOld} is
\begin{gather}
\mathcal{M}_{n,d}(\{\lambda_i,\tilde\lambda_i,\eta_i\})
= \int \frac{\prod\limits_{a=0}^d \mathrm{d}^{4|8}\mathcal{Z}_a}{{\rm vol\,GL(2;\mathbb{C })}} \frac{{\det}'(\Phi)
{\det}'(\widetilde{\Phi})}{|1\dots n|^2} \prod\limits_{i=1}^n (\sigma_i\mathrm{d}\sigma_i)h_i(\mathcal{Z}(\sigma_i)).
\label{N=8}
\end{gather}
In the rest of the paper we will use whichever form is more convenient for the argument at hand.

\section{BCFW recursion}

In this section, we will prove the conjecture of~\cite{Cachazo:2012kg} by showing that the $\mathcal{M}_{n,d}$ def\/ined by equation~\eqref{N=8}
correctly obeys BCFW recursion.
There are four aspects to the proof.
Firstly, we must show that the formula correctly reproduces the 3-particle amplitudes that seed the recursion.
This step is straightforward.
Next, we must show that under the BCFW shift\footnote{In Appendix~\ref{sec:determinants} we prove that $\mathcal{M}_{n,d}$ is completely
permutation symmetric in the external states.
Hence there is no loss of generality in considering this BCFW shift.}
\begin{gather}
\lambda_1\to\lambda_1+z\lambda_n
\qquad
\tilde\lambda_n\to\tilde\lambda_n-z\tilde\lambda_1,
\qquad
\chi_n\to\chi_n-z\chi_1
\label{shift}
\end{gather}
the integral in~\eqref{N=8} decays at least as fast as $1/z$ in the limit that $z\to\infty$.
Thirdly, we must show that $\mathcal{M}_{n,d}$ has a~pole whenever the sum of momenta of any two or~more particles becomes null, with the
residue of this pole being the product of two subamplitudes.
Finally, we complete the argument by showing that $\mathcal{M}_{n,d}$ has no poles other than the physical ones.
This being the case, the usual BCFW contour argument~\cite{Britto:2004ap} may be applied to construct $\mathcal{M}_{n,d}$ recursively from
smaller amplitudes.
Equation~\eqref{N=8} will then agree with the tree amplitudes in $\mathcal{N}=8$ supergravity since it satisf\/ies the same recursion relation
with the same initial conditions~\cite{ArkaniHamed:2008gz,ArkaniHamed:2008yf, Bedford:2005yy,Benincasa:2007qj,Cachazo:2005ca,Cheung:2008dn}.

In fact, it is known that gravitational scattering amplitudes decay as $1/z^2$ under the BCFW shift~\cite{ArkaniHamed:2008yf,Cheung:2008dn}.
We will see that $\mathcal{M}_{n,d}$ has precisely this behaviour quite transparently.
Although this fact can be shown using Lagrangian techniques~\cite{ArkaniHamed:2008yf,Cheung:2008dn}, the proof is rather opaque from the view
point of the $S$-matrix.

\subsection{3-particle seed amplitudes}

We f\/irst check that~\eqref{N=8} yields the correct 3-point amplitudes.
For the $\overline{\mathrm{MHV}}$ we have $n=3$ and~$d=0$, so that the map $\mathcal{Z}$ is constant, $\mathcal{Z}(\sigma)=\mathcal{Z}$.
We can remove all three rows and~columns of~$\Phi$ and~it is simple to show that ${\det}'(\Phi)$ cancels\footnote{Here and~below we take
${\det}'(\Phi)$ to be def\/ined as in~\eqref{detprimedef}; see the discussion after equation~\eqref{Rdenom}.} the factor of~$|123|^2$ in the
denominator of~\eqref{N=8}.
We can also remove two of the three rows and~columns from $\widetilde{\Phi}$.
Choosing these to be the f\/irst and~second rows and~the f\/irst and~third columns, the reduced determinant becomes
\begin{gather*}
{\det}'(\widetilde{\Phi}) = \frac{1}{(12)(23)(31)}\left[\frac{\partial}{\partial\mu(\sigma_2)}\frac{\partial}{\partial\mu(\sigma_3)}\right].
\end{gather*}
The denominator $(12)(23)(31)$ is exactly compensated by the Jacobian from f\/ixing worldsheet SL($2;\mathbb{C }$) invariance, so~\eqref{N=8}
reduces to
\begin{gather*}
\mathcal{M}_{3,0}= [23] \int \frac{\mathrm{d}^{4|8}\mathcal{Z}}{\rm vol(\mathbb{C }^*)} t_2t_3\prod\limits_{i=1}^3\frac{\mathrm{d}
t_i}{t_i^3} \bar\delta^2(\lambda_i-t_i\lambda)\exp(t_i\llbracket\mu\tilde\lambda_i\rrbracket)
\\
\phantom{\mathcal{M}_{3,0}}{}
 = [23]\int\prod\limits_{i=2}^3\frac{\mathrm{d}
t_i}{t_i^2}\bar\delta^2(\lambda_i-t_i\lambda_1)\bar\delta^{2|8}\big(\tilde\lambda_1+t_2\tilde\lambda_2+t_3\tilde\lambda_3\big),
\end{gather*}
where in going the second line we f\/ixed the $\mathbb{C }^*$ scaling by setting $t_1=1$ and~then performed the $\mathrm{d}^{4|8}\mathcal{Z}$
integral.
Using the two bosonic $\delta$-functions involving the $\tilde\lambda$'s to f\/ix $t_2$ and~$t_3$ shows that
\begin{gather*}
\mathcal{M}_{3,0} = \frac{\delta^4\left(\sum\limits_{i=1}^3
p_i\right)\delta^{0|8}\left(\eta_1[23]+\eta_2[31]+\eta_3[12]\right)}{([12][23][31])^2},
\end{gather*}
which is exactly the required 3-point $\overline{\mathrm{MHV}}$ amplitude.

For general wavefunctions $h_i$, $\mathcal{M}_{0,3}$ is simply the cubic vertex of the twistor action for self-dual $\mathcal{N}=8$
supergravity~\cite{Mason:2007ct}
\begin{gather}
S_{\rm sdG} = \int_{\mathbb{PT}} \mathrm{D}^{3|8}\mathcal{Z} \wedge\left( h\wedge\bar{\partial} h+\frac{2}{3}h\wedge\left\{h,h\right\}\right),
\label{sdGact}
\end{gather}
evaluated on on-shell states.
In this action,
\begin{gather*}
\left\{f,g\right\}:=I^{\rm ab}\frac{\partial f}{\partial Z^{\rm a}}\frac{\partial g}{\partial Z^{\rm b}} = \left[\frac{\partial
f}{\partial\mu}\frac{\partial g}{\partial\mu}\right]
\end{gather*}
is the Poisson bracket associated to the inf\/inity twistor.
Notice that the negative homogeneity of the Poisson bracket ensures the interaction term scales the same way as the kinetic term, and~that each
balances the scaling of the $\mathcal{N}=8$ measure.

The linearized f\/ield equations of~\eqref{sdGact} state that $h(Z)$ represents a~class in $H^{0,1}(\mathbb{PT}, \mathcal{O}(2))$, as required by
the Penrose transform for massless free f\/ields of helicity~$+2$.
At the non-linear level, the f\/ield equations state that $\bar\partial + \{h,\cdot\}$ def\/ines an integrable almost complex structure on~$\mathbb{PT}$ that is compatible with the Poisson structure.
This is exactly the content of Penrose's non-linear graviton construction~\cite{Penrose:1976jq}.
A~twistor space with an integrable almost complex structure corresponds to a~conformal equivalence class of space-times with self-dual Weyl
tensor.
The additional information that the complex structure is compatible with the Poisson structure picks a~distinguished metric in the conformal
class that satisf\/ies the vacuum Einstein equations.
The presence of the inf\/inity twistor in $\mathcal{M}_{\overline{\mathrm{MHV}}}$ is thus a~direct consequence of its presence in the self-dual
action, ref\/lecting the very nature of the non-linear graviton construction.

For the 3-point MHV we have $(n,d)=(3,1)$, so that $\mathcal{Z}(\sigma) = A\sigma^{\underline 0} + B\sigma^{\underline 1}$.
We can now remove two rows and~columns from~$\Phi$.
Choosing these to be the f\/irst and~second rows and~the f\/irst and~third columns, equation~\eqref{detprimestandard} gives
\begin{gather*}
\frac{{\det}'(\Phi)}{|123|^2} = \frac{\langle23\rangle}{(23)}\frac{1}{t_2t_3} = \langle AB\rangle,
\end{gather*}
where the second equality holds on the support of the $\delta$-functions for $\lambda_i$.
We can remove all three rows and~columns from $\widetilde{\Phi}$, so
\begin{gather*}
{\det}'(\widetilde{\Phi}) = \frac{1}{(12)^2(23)^2(31)^2}=\frac{\langle
AB\rangle^6}{\langle12\rangle^2\langle23\rangle^2\langle31\rangle^2}\prod\limits_{i=1}^3  t_i^4.
\end{gather*}
The integral $\mathcal{M}_{3,1}$ then becomes
\begin{gather*}
\mathcal{M}_{3,1} = \frac{1}{\langle12\rangle^2\langle23\rangle^2\langle31\rangle^2} \\
\hphantom{\mathcal{M}_{3,1} =}{}
\times \int\frac{\mathrm{d}^{4|8}A\mathrm{d}^{4|8}B}{\rm vol
\,GL(2)} \langle{AB}\rangle^7 \prod\limits_{i=1}^3(\sigma_i\mathrm{d}\sigma_i)\mathrm{d} t_i\, t_i  \bar\delta^2(\lambda_i - t_i
\lambda(\sigma_i))\exp\big(t_i\llbracket\mu(\sigma_i)\tilde\lambda_i\rrbracket\big).
\end{gather*}
The integrals over $t_i$ and~$\sigma_i$ integrals may be f\/ixed by the $\delta$-functions.
The integrals over $(\mu_{A,B},\chi_{A,B})$ then provide a~super-momentum conserving $\delta$-function, while the four remaining integrals over
$|A\rangle$ and~$|B\rangle$ are compensated by the GL(2).
Overall, we have
\begin{gather*}
\mathcal{M}_{3,1} = \frac{\delta^{4|16}\left(\sum\limits_i p_i\right)}{\langle12\rangle^2\langle23\rangle^2\langle31\rangle^2},
\end{gather*}
exactly the $3$-particle MHV amplitude in $\mathcal{N}=8$ supergravity, as expected.

The 3-particle amplitudes that seed BCFW recursion are thus correctly reproduced by the integral~\eqref{N=8}.

\subsection[Decay as $z\to\infty$]{Decay as $\boldsymbol{z\to\infty}$}

We now investigate the behaviour of~$\mathcal{M}_{n,d}$ under the BCFW shift in the limit that the shift parameter $z\to\infty$.
We shall see that the highly non-trivial fact that the gravitational amplitudes decay as $1/z^2$ in this limit is made manifest by~\eqref{N=8}.

At degree~$d$, we can remove $d+2$ rows and~columns from $\widetilde{\Phi}$ and~$n-d$ rows and~columns from~$\Phi$.
Hence, since we are only interested in BCFW recursion for $1\leq d\leq n-3$, we can always remove at least two rows and~columns from each.
With the shift~\eqref{shift} that af\/fects only~$|1\rangle$ and~$|n]$, we choose the removed rows and~columns to include 1 and~$n$ in both cases.
In addition, we choose one of the arbitrary points $p_r\in\Sigma$ in~\eqref{phidef}
to be $\sigma_1$ and~another to be $\sigma_n$ so that the terms $j=1,n$
drop out of the sum over~$j$ the diagonal elements $\Phi_{ii}$.
Similarly, we choose the arbitrary points $p_a\in\Sigma$ in~\eqref{tphidef} to include $\sigma_1$ and~$\sigma_n$.
With these choices, the external data~$|1\rangle$,~$|n\rangle$ and~$|1]$,~$|n]$ does not appear in the determinants ${\det}'(\Phi){\det}'(\widetilde{\Phi})$.
Thus, after integrating out the $(\mu,\chi)$ components of the map, the shift~\eqref{shift} af\/fects~\eqref{N=8} only by changing the arguments
of the $\delta$-functions involving $\lambda_1$ and~$\tilde\lambda_n$.
On the coordinate patch $\sigma^{\underline\alpha}=(1,u)$ of the worldsheet, these shifted $\delta$-functions become
\begin{gather*}
\bar\delta^2\left(\lambda_1+z\lambda_n- t_1\sum\limits_{a=0}^d\rho_a u_1^a\right)
\prod\limits_{a=0}^d\bar\delta^{2|8}\left(\sum\limits_{j=1}^n t_j\tilde\lambda_j u_j^a-zt_n\tilde\lambda_1u_n^a\right),
\end{gather*}
where $\rho_a$ is the $\lambda$ part of the map $\mathcal{Z}$.

To absorb these shifts, we introduce new worldsheet variables $(\hat u,\hat t)$ for particle 1, def\/ined by
\begin{gather}
\hat t_1 := t_1 - zt_n,
\qquad
\hat t_1 (\hat{u}_1)^d := t_1 (u_1)^d - zt_n(u_n)^d,
\label{hatvarsdef}
\end{gather}
where~$z$ is the BCFW shift parameter.
This def\/inition absorbs the shifts in the $\delta$-functions, up to terms that vanish as $z\to\infty$.
Specif\/ically, the argument of the shifted $\lambda_1$ $\delta$-function becomes
\begin{gather*}
\lambda_1-\hat{t}_1\sum\limits_{a=0}^d\rho_a\left(\frac{d-a}{d}u_n^a+\frac{a}{d}\hat{u}_1^du_n^{a-d}\right) + \mathcal{O}(1/z)
\end{gather*}
while the arguments of the $\delta$-functions involving the $\tilde\lambda$'s become
\begin{gather*}
\sum\limits_{j=2}^nt_j\tilde\lambda_j u_j^a + \hat{t}_1\tilde\lambda_1 \left(\frac{d-a}{d}u_n^a+\frac{a}{d}\hat{u}_1^du_n^{a-d}\right) +
\mathcal{O}(1/z).
\end{gather*}
The important point is that the new variables $(\hat{u}_1,\hat{t}_1)$ remain f\/inite as $z\to\infty$.
Therefore, to study the behaviour of~\eqref{N=8} in this limit, we should express its integrand in terms of these variables.

Begin with the measure for particle 1.
It follows from~\eqref{hatvarsdef} that
\begin{gather*}
(\sigma_1\mathrm{d}\sigma_1) = \mathrm{d} u_1 = \left(\frac{zt_nu_n^d+\hat{t}_1{\hat u}_1^d}{zt_n+\hat{t}_1}\right)^{\frac{1}{d}-1}
\ \stackrel{z\to\infty}\longrightarrow \ \frac{\hat{t}_1}{zt_n} u_n^{\frac{d-1}{d}}\hat{u}_1^{d-1}\mathrm{d} \hat{u}_1,
\\
\frac{\mathrm{d} t_1}{t_1^3} = \frac{\mathrm{d}\hat{t}_1}{(zt_n+\hat{t}_1)^3} \ \stackrel{z\to\infty}\longrightarrow \
\frac{1}{z^3}\frac{\mathrm{d}\hat{t}_1}{t_n^3},
\end{gather*}
where we have dropped terms that wedge to zero against the measure for particle~$n$.
Thus the integration measure of~\eqref{N=8} falls as $z^{-4}$ as the shift parameter tends to inf\/inity.
Similarly, we f\/ind that
\begin{gather*}
(1j) = u_1-u_j = \left(\frac{zt_nu_n^d+\hat{u}_1^d\hat{t}_1}{zt_n+\hat{t}_1}\right)^{\frac{1}{d}}-u_j
\
\stackrel{z\to\infty}{\longrightarrow}
\
(nj)
\end{gather*}
whenever $(nj)\neq 0$, but that
\begin{gather*}
(1n) = \left(\frac{zt_nu_n^d+\hat{u}_1^d\hat{t}_1}{zt_n+\hat{t}_1}\right)^{\frac{1}{d}}-u_j
\
\stackrel{z\to\infty}\longrightarrow
\
\frac{u_n\hat{t}_1}{z t_n}\left(\frac{\hat{u}_1^d}{u_n^d}+1\right),
\end{gather*}
so that the special case of~$(1n)$ decays as $z^{-1}$, as the order $z^0$ term cancels exactly.

We now investigate the occurrence of~$(1n)$ and~$t_1$ in the integrand of~$\mathcal{M}_{n,d}$, since these are the only terms that have
non-trivial large~$z$ behaviour.
We can always choose to remove row and~columns 1 and~$n$ from both~$\Phi$ and~$\widetilde{\Phi}$.
This does not quite suf\/f\/ice to remove $(1n)$ and~$t_1$ from the matrices, because they still appear in the diagonal terms
\begin{gather*}
\Phi_{ii} = -\sum\limits_{j\neq i} \frac{\langle ij\rangle}{(ij)}\frac{1}{t_it_j} \prod\limits_{r=0}^{\tilde
d}\frac{(jp_r)}{(ip_r)}\frac{\prod\limits_{k\neq i}(ik)}{\prod\limits_{l\neq j}(jl)},
\qquad
\widetilde{\Phi}_{ii} = -\sum\limits_{j\neq i} \frac{[ij]}{(ij)}t_it_j\prod\limits_{a=0}^{d}\frac{(jp_a)}{(ip_a)}.
\end{gather*}
By further choosing one of the $p_r$ and~one of the $p_a$ to be $\sigma_1$ the summand with $j=1$ vanishes in each of these matrices, and~since
$i\neq1,n$ the matrices themselves approach a~constant value as $z\to\infty$, obtained by simply replacing $\sigma_1\to\sigma_n$.

Aside from the measure then, the only pieces of the integrand which af\/fect the large~$z$ behaviour are the Vandermonde determinants in the
def\/inition of the reduced determinants.
Since we have removed rows and~columns 1 and~$n$, the Vandermonde determinants associated with~$\Phi$ are independent of\footnote{We recall that
this determinant can be def\/ined to absorb the overall factor of~$|1\dots n|^{-2}$ in~\eqref{N=8}, whereupon the associated Vandermonde
determinants are those of the rows and~columns that {\it remain} in~$\Phi$.} $(1n)$.
However, we f\/ind that the def\/inition of~${\det}'(\widetilde{\Phi})$ involves the denominator
\begin{gather*}
|1n r_1\dots r_d|^2 \propto (1n)^2,
\end{gather*}
where $r_1,\ldots, r_d$ are the other rows and~columns that were removed from $\widetilde{\Phi}$.
This factor, appearing in the denominator of~${\det'}(\widetilde{\Phi})$, behaves as $1/z^2$ in the large shift limit.
Combined with the $1/z^4$ behaviour of the integration measure, we see that $\mathcal{M}_{n,d}(z)\propto 1/z^2$ as $z\to\infty$.
This ensures that the BCFW integrand $ \mathcal{M}_{n,d}(z)\mathrm{d} z/z$ has no pole at inf\/inity, allowing the BCFW residue theorem to
proceed.
It is quite remarkable that the formula~\eqref{N=8} for $\mathcal{M}_{n,d}$ reproduces the correct $1/z^2$ behaviour of gravity so
transparently.
We repeat that this behaviour is highly non-trivial to prove by any other means, and~yet is a~key property of gravitational scattering
amplitudes~\cite{Spradlin:2008bu}.

Incidentally, exactly the same argument as above may be applied to the Witten-RSV formula~\cite{Roiban:2004yf,Witten:2003nn}
\begin{gather}
\mathcal{A}_{n,d} = \int\frac{\prod\limits_{a=0}^d \mathrm{d}^{4|4}\mathcal{Z}_a}{{\rm vol\,GL(2;\mathbb{C })}}\prod\limits_{i=1}^n
\frac{(\sigma_i\mathrm{d}\sigma_i)}{(\sigma_i\sigma_{i+1})} \frac{\mathrm{d} t_i}{t_i} \delta^2(\lambda_i-t_i\lambda(\sigma_i))
\exp\big(t_i\llbracket \mu(\sigma_i)\tilde\lambda_i\rrbracket\big)
\label{N=4}
\end{gather}
for tree-level scattering amplitudes in $\mathcal{N}=4$ SYM.
We now have
\begin{gather*}
\frac{\mathrm{d} t_1}{t_1} \mathrm{d} u_1\stackrel{z\to\infty}\longrightarrow \frac{1}{z^2}\frac{\hat{t}_1\mathrm{d}\hat{t}_1}{t_n^2}
u_n^{\frac{d-1}{d}}\hat{u}_1^{d-1}\mathrm{d} \hat{u}_1
\end{gather*}
so that the measure decays as $1/z^2$, while $(1n)$ still behaves as $1/z$ under~\eqref{hatvarsdef} so that the decay is softened to $1/z$
overall.
Had we chosen to shift external particles that were not adjacent in the colour ordering, all the $(i\,i+1)$ brackets would have approached
constants as $z\to\infty$.
Representing Yang--Mills amplitudes by $\mathcal{A}_{n,d}$ thus makes it manifest that they behave as $1/z^2$ under BCFW shifts of non-adjacent
particles~\cite{ArkaniHamed:2008yf,Cheung:2008dn}.
Once again, this fact is very dif\/f\/icult to see by any other means except the Grassmanian formulation of Yang--Mills
amplitudes~\cite{ArkaniHamed:2009dn,Mason:2009qx}.

\subsection{Multi-particle factorization}

The main ingredient in the proof is multiparticle factorization\footnote{Factorization properties of the form~\eqref{N=4} of Yang--Mills
amplitudes were investigated in~\cite{Gukov:2004ei,Skinner:2010cz, Vergu:2006np} and~the discussions there are closely related to the argument
here.}.
Gravitational tree amplitudes have a~pole whenever the sum~$P$ of any two or~more external momenta becomes null, and~the residue of this pole is
the product of two subamplitudes, summed over the helicities of the particle being exchanged.
More specif\/ically, divide the particles into two sets~$L$ and~$R$ and~call
\begin{gather*}
P_L:=\sum\limits_{i\in L} p_i,
\qquad
P_R:=\sum\limits_{j\in R} p_j.
\end{gather*}
Then the amplitude behaves as
\begin{gather}
\mathcal{M}(\Lambda_1,\ldots,\Lambda_n) \stackrel{P_L^2\to0}\longrightarrow \delta^4\left(\sum\limits_{k=1}^n p_k\right)
\int\mathrm{d}^8\eta\widetilde{\mathcal{M}}_{L}(\{\Lambda_i\},\Lambda)\frac{1}{P_L^2} \widetilde{\mathcal{M}}_{R}(-\Lambda,\{\Lambda_j\})+\cdots,
\label{factorprod}
\end{gather}
where $\Lambda_i=\{\lambda_i,\tilde\lambda_i,\eta_i\}$ is shorthand for the spinor momenta, and~$\Lambda$ represents the spinor momenta of the
internal particle in the strict limit that $P_L^2=0$.
In this equation, $\widetilde{\mathcal{M}}$ represents an amplitude stripped of its overall (bosonic) momentum $\delta$-function.
We can restore these $\delta$-functions by writing
\begin{gather}
\delta^4\left(\sum\limits_{k=1}^n p_k\right) \int\mathrm{d}^8\eta \widetilde{\mathcal{M}}_{L}(\{\Lambda_i\},\Lambda)\frac{1}{P_L^2}
\widetilde{\mathcal{M}}_{R}(-\Lambda,\{\Lambda_j\})
\nonumber
\\
=\int\frac{\mathrm{d}^4p}{p^2}\delta^4\left(P_L + p\right)\widetilde{\mathcal{M}}_{L}(\{\Lambda_i\},\Lambda)
\delta^4\left(-p+P_R\right)\widetilde{\mathcal{M}}_{R}(-\Lambda,\{\Lambda_j\})
\label{deltarestore}
\\
=\int\langle\lambda\mathrm{d}\lambda\rangle\mathrm{d}^{2|8}\tilde\lambda\frac{\mathrm{d} s^2}{s^2}
\delta^4\big(P_L+\lambda\tilde\lambda + s^2q\big)\widetilde{\mathcal{M}}_{L}(\{\Lambda_i\},\Lambda)
\delta^4\big(-\lambda\tilde\lambda-s^2q+P_R\big)
\widetilde{\mathcal{M}}_{R}(-\Lambda,\{\Lambda_j\}),
\nonumber
\end{gather}
where in the last line we have parameterized~$p$ by
\begin{gather}
p = \lambda\tilde\lambda + s^2 q,
\label{pas2null}
\end{gather}
where $\lambda\tilde\lambda$ and~$q$ are null momenta, with~$q$ f\/ixed, and~$s^2$ is a~scalar parameter chosen for later convenience.
Any 4-momentum may be parametrized this way.
Notice also that $p^2=P_L^2 = s^2\langle\lambda|q|\tilde\lambda]$.

Suppose we approach the factorization channel by taking the limit as $s^2\to 0$.
If we wish to recover the amplitude~\eqref{factorprod} then the $\mathrm{d}^4p$ integral in the second line of~\eqref{deltarestore} should be
performed over a~copy of real momentum space.
However, as the amplitude itself is diverging, it is more sensible to compute the residue of the pole.
This may be done by changing the contour in the f\/inal line to be an $S^1$ encircling the pole at $s^2=0$, together with an integral over the
on-shell phase space of the intermediate particle.
One f\/inds
\begin{gather}
\mathop{\operatorname{Res}}\limits_{P_L^2=0}\mathcal{M}(\Lambda_1,\ldots,\Lambda_n) =
\int\langle\lambda\mathrm{d}\lambda\rangle\mathrm{d}^{2|8}\tilde\lambda\mathcal{M}_{L}(\{\Lambda_i\},\Lambda)\mathcal{M}_{R}(-\Lambda,\{\Lambda_j\}),
\label{factorres}
\end{gather}
where the $\delta$-functions are now naturally incorporated into the subamplitudes.
In particular, this formula shows that the residue itself has no memory of the direction in which the factorization channel was approached.

\begin{figure}[t]\centering
$\mathop{\operatorname{Res}}\limits_{s^2=0}\mathcal{M}(\mathcal{Z}_1,\ldots,\mathcal{Z}_n) = \displaystyle{\int}\mathrm{D}^{3|8}\mathcal{Z}
$\raisebox{-10mm}{\includegraphics[height=25mm]{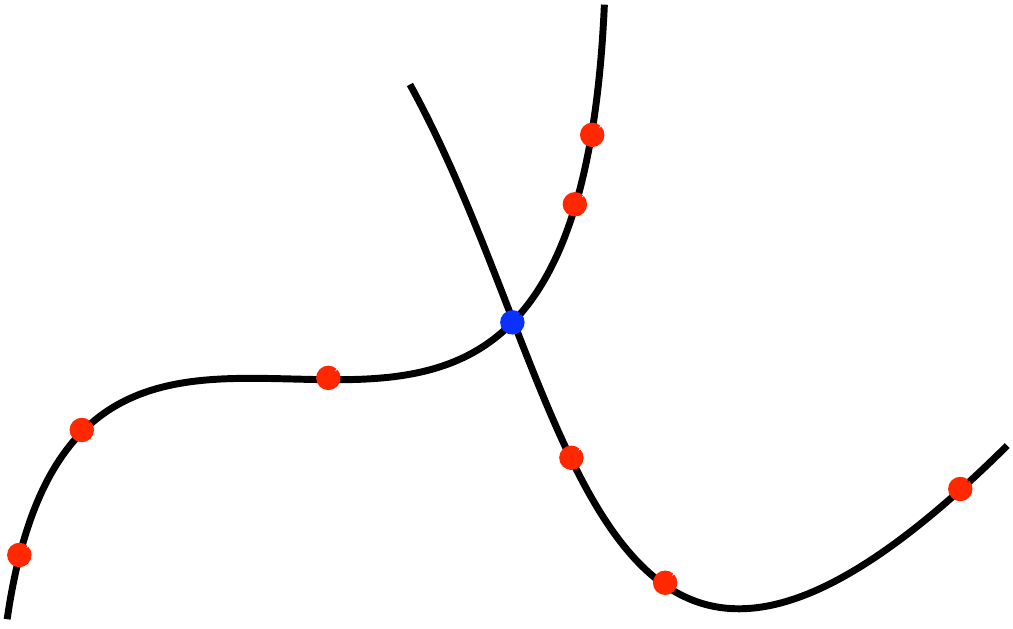}}
\caption{On twistor space, the residue of a~factorization channel looks like a~nodal curve with the location $\mathcal{Z}$ of the node
integrated over.}
\label{fig:nodal}
\end{figure}

We must show that $\mathcal{M}_{n,d}$ in~\eqref{N=8} has the same property.
It will actually be convenient f\/irst to rewrite the residue on twistor space by transforming the external and~internal $\Lambda$'s to twistor
variables\footnote{This is implemented by use of wave functions $h_i=\delta^{3|8}(Z,Z_i):=\int \delta^{4|8}(Z_i-tZ) d t/t^3$ in~\eqref{N=8}
instead of momentum eigenstates; in split signature this is the half-Fourier transform of the momentum-space version, see discussion
around~\eqref{twistoreig}.}.
Doing so,~\eqref{factorres} becomes~\cite{ArkaniHamed:2009si,Bullimore:2009cb,Mason:2009sa,Skinner:2010cz}\footnote{We do not distinguish the
symbol $\mathcal{M}$ for momentum space amplitudes from that of twistor space ones.
Which is meant should be clear from the context.}
\begin{gather}
\mathop{\operatorname{Res}}_{s^2=0}\mathcal{M}(\mathcal{Z}_1,\ldots,\mathcal{Z}_n) = \int \mathrm{D}^{3|8}\mathcal{Z}\wedge\mathcal{M}_L(\{\mathcal{Z}_i\},
Z)\wedge\mathcal{M}_R(Z,\{\mathcal{Z}_j\}),
\label{twistorres}
\end{gather}
where $\{\mathcal{Z}_i\}$ and~$\{\mathcal{Z}_j\}$ are the sets of twistors associated with external states on the~$L$ and~$R$ subamplitudes.
Notice that on twistor space, $\mathcal{N}=8$ gravitational amplitudes are homogeneous of degree $+2$ in each of their arguments.
Under the assumption (valid at least for 3-particle amplitudes) that these gravitational subamplitudes are associated with curves in twistor
space, we see that the residue on a~factorization channel corresponds to a~nodal curve, with the location~$\mathcal{Z}$ of the node integrated
over the space (see Fig.~\ref{fig:nodal}). Therefore, to prove that~$\mathcal{M}_{n,d}$ as given by~\eqref{N=8} obeys BCFW recursion~-- and~therefore agrees
with all tree amplitudes in $\mathcal{N}=8$ supergravity~-- we must show both that it has a~simple pole on the boundary of the moduli space
where the curve degenerates, and~further that the residue of this pole is given by~\eqref{twistorres}.

A standard way to describe the decomposition of a~rational curve into a~nodal curve is introduce a~complex parameter~$s$ and~model the rational
curve as the conic\footnote{We emphasize that this is a~model for the abstract worldsheet $\Sigma$ before it is mapped to twistor space.}
\begin{gather*}
\Sigma_s = \big\{xy=s^2z^2\big\} \subset\mathbb{CP}^2,
\end{gather*}
where $(x,y,z)$ are homogeneous coordinates on the complex projective plane.
The homogeneous coordinates $\sigma^{\underline\alpha}=(\sigma^{\underline0},\sigma^{\underline1})$ intrinsic to the $\mathbb{CP}^1$ worldsheet
are related to these coordinates by
\begin{gather*}
(x,y,z)=\left(\big(\sigma^{\underline 0}\big)^2,\big(\sigma^{\underline1}\big)^2,\frac{\sigma^{\underline0}\sigma^{\underline1}}{s}\right).
\end{gather*}
The degeneration of the curve is controlled by the parameter $s^2$, which we will show is the same parameter as appears in~\eqref{pas2null}.
In the limit we have
\begin{gather*}
\lim_{s\to0}\Sigma_s = \Sigma_L\cap\Sigma_R,
\end{gather*}
where the $\mathbb{CP}^1$'s $\Sigma_L$ and~$\Sigma_R$ are def\/ined by
\begin{gather*}
\Sigma_L = \{y=0\}\subset\mathbb{CP}^2,
\qquad
\Sigma_R = \{x=0\}\subset\mathbb{CP}^2
\end{gather*}
so that $(z,x)$ form homogeneous coordinates on $\Sigma_L$ and~$(z,y)$ form homogeneous coordinates on $\Sigma_R$.
The good homogeneous coordinates intrinsic to $\Sigma_{L,R}$ are therefore
\begin{gather*}
\sigma_L^{\underline\alpha} =(z,x) =\sigma^{\underline0}\left(\sigma^{\underline1}/s,\sigma^{\underline0}\right),
\qquad
\sigma_R^{\underline\alpha} =(z,y)=\sigma^{\underline1}\left(\sigma^{\underline0}/s,\sigma^{\underline1} \right)
\end{gather*}
and~the af\/f\/ine coordinate $u=\sigma^{\underline 1}/\sigma^{\underline0}$ on $\Sigma_s$ is related to the af\/f\/ine coordinates $u_{L,R}$ on
$\Sigma_{L,R}$~by
\begin{gather}
u_L=\frac{s}{u},
\qquad
u_R=s u.
\label{affinecoords}
\end{gather}
With this choice of coordinates, the node $\Sigma_L\cap\Sigma_R$ is the point $x=y=0\in\mathbb{CP}^2$ and~is also at the origin in each of the
af\/f\/ine coordinates $u_{L,R}$.

As the curve degenerates, the~$n$ marked points distribute themselves among the component curves $\Sigma_{L,R}$, with at least two of these
points on each curve component.
In the degeneration limit, any such distribution def\/ines a~boundary divisor in the moduli space $\overline M_{0,n}$ of $n$-pointed rational
curves, with the locations of the marked points considered up to SL$(2;\mathbb{C })$ transformations.
The parameter $s^2$ is then a~coordinate transverse to this boundary divisor, which lies at $s^2=0$.

Ordinarily, we think of coordinates on $\overline M_{0,n}$ as given by a~choice of~$n-3$ independent cross-ratios of the marked points.
No choice of these cross ratios provides coordinates globally on~$\overline M_{0,n}$, but we can always make a~choice such that a~particular
boundary divisor arises when one or~more cross-ratios approach zero, so that in some conformal frame the marked points in the numerator of these
cross ratios are colliding.
To relate this description to $s^2$, consider the cross-ratios
\begin{gather}
x_k:= \frac{(1k)(n-1\, n)}{(n1)(k\,n-1)},
\label{xratio}
\end{gather}
where without loss of generality we assume that our boundary divisor has $1\in L$ and~$n-1$, $n\in R$.
To study the degeneration, marked points should be described in terms of the coordinates $u_L$ or~$u_R$ as appropriate.
Using
\begin{gather}
u_i-u_j =
\begin{cases}
s\displaystyle{\frac{u_{jL}-u_{iL}}{u_{iL}u_{jL}}},
&i,j\in L,
\vspace{1mm}\\
\displaystyle{\frac{u_{iR}-u_{jR}}{s}},
&i,j\in R,
\vspace{1mm}\\
\displaystyle{\frac{s^2-u_{iL}u_{jR}}{su_{jR}}},
&i\in L,
\quad
j\in R
\end{cases}
\label{(ij)LR}
\end{gather}
we see that
\begin{gather*}
x_i = s^2 \frac{(u_{1L}-u_{iL})(u_{n-1 R}-u_{nR})}{u_{1L}u_{iL} u_{nR}u_{n-1 R}} + \mathcal{O}\big(s^4\big)
\end{gather*}
when $i\in L$, whereas
\begin{gather*}
x_j = \frac{u_j(u_{n-1 R}-u_{nR})}{u_{nR}(u_{jR}-u_{n-1R})} + \mathcal{O}\big(s^2\big)
\end{gather*}
when $j\in R$.
Consequently, as we approach the boundary divisor, any ratio~$x_i/x_j$ with $i\in L$ and~$j\in R$ will vanish as~$s^2$, whereas any such ratio
with~$i$ and~$j$ limiting onto the same curve components remains f\/inite, provided we approach a~generic point of the boundary divisor
(i.e., we only consider a~single degeneration).
We can now extract $s^2$ by def\/ining rescaled cross-ratios~$y_i$ by
\begin{gather*}
x_i =: s^2 y_i
\qquad
\text{for}
\quad
i\in L,
\end{gather*}
where the $y_i$ are to be considered only up to an overall scaling.

The factor
\begin{gather*}
\mathrm{d}\mu:=\frac{1}{\rm vol(SL(2;\mathbb{C }))} \prod\limits_{i=1}^n (\sigma_i\mathrm{d} \sigma_i)
\end{gather*}
provides meromorphic top form on the moduli space.
This form cannot be written in terms of the cross-ratios alone since it has non-zero homogeneity in each of the $\sigma$s.
However, f\/ixing the SL$(2:\mathbb{C })$ by freezing $1$, $n-1$ and~$n$, at least locally we can write
\begin{gather}
\frac{1}{\rm vol(SL(2;\mathbb{C }))} \prod\limits_{i=1}^n (\sigma_i\mathrm{d} \sigma_i)= f(u_i)
\prod\limits_{i=2}^{n-2}\mathrm{d} x_i = f(u_i)
\prod\limits_{\substack{i\in L\\i\neq 1}}\mathrm{d} x_i \prod\limits_{\substack{j\in R
\\
j\neq n-1,n}} \mathrm{d} x_j
\label{gaugefixed}
\end{gather}
in terms of~$n-3$ of the cross-ratios~\eqref{xratio} and~where the function\footnote{More precisely, $\frac{1}{{\rm vol}({\rm SL}(2;\mathbb{C
}))}\prod (\sigma_i\mathrm{d}\sigma_i)$ is an $n-3$ form with values in $\otimes_i\mathbb{L}_i^{-1}$, where $\mathbb{L}_i$ is the tautological
line bundle on $\overline M_{0,n}$ whose f\/ibre is the holomorphic cotangent space of~$\Sigma$ at~$i$.
The function $f(u_i)$ is thus really a~section of~$\otimes_i\mathbb{L}_i^{-1}$.}
\begin{gather*}
f(u_i):= \left[\prod\limits_{i\neq n-1}(i\,n-1)^2\right]\left[\frac{(1n)}{(n-1\,n)(1\,n-1)}\right]^{n-2}
\end{gather*}
absorbs the homogeneity.

Upon transforming to the new coordinates, for $i\in L$ we wish to replace the $x_i$ cross-ratios by $s^2$ and~$y_i$, with the $y_i$ treated
projectively.
To leading order in $s^2$, the measure for these~$L$ cross-ratios becomes
\begin{gather*}
\prod\limits_{\substack{i\in L\\i\neq1}}
\mathrm{d} x_i = s^{2(n_L-2)}\mathrm{d} s^2\epsilon^{i_1i_2\cdots i_{n_L-1}} y_{i_1}\mathrm{d} y_{i_2}\cdots \mathrm{d} y_{i_{n_L-1}}
=s^{2(n_L-2)}\mathrm{d} s^2 y_2 \prod\limits_{\substack{i\in L\\i\neq 1,2}} \mathrm{d} y_i,
\end{gather*}
where in the second line we work on the af\/f\/ine patch $y_2={\rm const}$.
The measure for the~$R$ cross-ratios is~$s$-independent and~stays as in~\eqref{gaugefixed}.
To put this measure in a~more familiar form, we now undo the transformation~\eqref{gaugefixed} separately on the left and~on the right, f\/inding
\begin{gather*}
\prod\limits_{\substack{i\in L\\i\neq 1,2}}\mathrm{d} y_i = \left[\frac{(u_{n-1R}-u_{nR})}{u_{n-1R}u_{nR}}\right]^{n_L-2}
\prod\limits_{\substack{i\in L\\i\neq 1,2}}\frac{\mathrm{d} u_{iL}}{u_{iL}^2},
\\
\prod\limits_{\substack{j\in R\\j\neq n-1,n}}\mathrm{d} x_j
=\left[(u_{n-1R}-u_{nR})\frac{u_{n-1R}}{u_{nR}}\right]^{n_R-2}
\prod\limits_{\substack{j\in R\\j\neq n-1,n}}\frac{\mathrm{d} u_{jR}}{(u_{jR}-u_{n-1R})^2}.
\end{gather*}
Putting all the pieces together, we have shown that in a~neighbourhood of the boundary divisor def\/ined by $s^2=0$, the measure for the
integration over marked points may be written as
\begin{gather}
\mathrm{d}\mu = s^{n_L-n_R -4}\mathrm{d}s^2\left[\frac{1}{{\rm vol}({\rm SL}(2;\mathbb{C }))}\prod\limits_{i\in L\cup\bullet}\mathrm{d}u_{iL}\right]
\left[\frac{1}{{\rm vol}({\rm SL}(2;\mathbb{C }))}\prod\limits_{j\in R\cup\bullet}\mathrm{d}
u_{jR}\right]\times\prod\limits_{i\in L}\frac{1}{u_{iL}^2}
\nonumber
\\
\phantom{\mathrm{d}\mu}
:= s^{n_L-n_R -4}\mathrm{d} s^2\mathrm{d}\mu_L\mathrm{d}\mu_R \times\prod\limits_{i\in L}\frac{1}{u_{iL}^2},
\label{measurewiths}
\end{gather}
where we recall that with the coordinates~\eqref{affinecoords} assumed a~gauge f\/ixing in which the node was at the origin in each curve
component.
Bearing in mind that the boundary divisor is naturally the product $\overline M_{0,n_L+1}\times \overline M_{0,n_R+1}$ of the moduli spaces for
rational curves with fewer marked points, the factors in square brackets are precisely the expected $(n_{L,R}+1) - 3$ forms on these spaces.
The form $\mathrm{d} s^2$ is normal to this boundary divisor.
Thus, to f\/ind the residue of our proposed gravity amplitude $\mathcal{M}_{n,d}$ in a~factorization channel, we must interpret the contour
in~\eqref{N=8} to include an $S^1$ factor that encircles the boundary divisor $s^2=0$, and~use this contour to compute the $\mathrm{d} s^2$
integral.
Of course, until we study the rest of the integrand in $\mathcal{M}_{n,d}$, it is not clear that we actually have a~simple pole there.

Armed with this description of a~neighbourhood of the factorization channel, we now investigate the behaviour of the rest of the integrand.
Begin by considering the map $\mathcal{Z}:\Sigma_s\to\mathbb{PT}$.
It is useful to pull out a~factor of~$u^{d_L}$ and~write
\begin{gather}
\mathcal{Z}(u,s) =u^{d_L}\left(\sum\limits_{a=1}^{d_L}\mathcal{Z}'_{d_L-a} u^{-c} + \mathcal{Z}_\bullet+
\sum\limits_{b=1}^{d_R}\mathcal{Z}'_{d_L+b} u^b\right)
 =u^{d_L}\left(\sum\limits_{a=1}^{d_L}\mathcal{Z}_a u_L^a + \mathcal{Z}_\bullet+ \sum\limits_{b=1}^{d_R}\mathcal{Z}_b
\frac{s^{2b}}{u_L^b}\right)
\nonumber
\\
\phantom{\mathcal{Z}(u,s)}
 =u^{d_L}\left(\sum\limits_{a=1}^{d_L}\mathcal{Z}_a \frac{s^{2a}}{u_R^a}+ \mathcal{Z}_\bullet +\sum\limits_{b=1}^{d_R}\mathcal{Z}_b
u_R^b\right)
\label{LRmaps}
\end{gather}
with the second or~third lines the appropriate description for particles limiting onto $\Sigma_{L,R}$, respectively.
The coef\/f\/icients $\mathcal{Z}_\bullet$, $\mathcal{Z}_a$ and~$\mathcal{Z}_b$ are related to the original map coef\/f\/icients $\mathcal{Z}'_c$ by
\begin{gather}
\mathcal{Z}_a:= s^a\mathcal{Z}'_{d_L-a},
\qquad
\mathcal{Z}_\bullet:= \mathcal{Z}'_{d_L},
\qquad
\mathcal{Z}_b:= s^b\mathcal{Z}'_{d_L+b}.
\label{rescaledZ}
\end{gather}
This shows that as $s^2\rightarrow 0$, the twistor curve $\mathcal{Z}(\Sigma)$ degenerates into a~pair of curves $\mathcal{Z}(\Sigma_L)$
and~$\mathcal{Z}(\Sigma_R)$ that are the images of degree $d_{L,R}$ maps, respectively, where\footnote{By forgetting the data of the map, the
moduli space $\overline M_{0,n}(\mathbb{PT},d)$ of degree~$d$ rational maps from an~$n$-pointed curve to $\mathbb{PT}$ admits a~morphism to
$\overline M_{0,n}$.
As we see in the text, a~boundary divisor in $\overline M_{0,n}(\mathbb{PT},d)$ is specif\/ied by a~boundary divisor in $\overline M_{0,n}$,
together with choices of degree $d_L$ and~degree $d_R$ maps on the two curve components, with $d_L+d_R=d$.} $d_L+d_R=d$.

As shown in~\cite{Vergu:2006np}, the $\delta$-functions involving $\lambda_i$ that are already present in the external wavefunctions combine
with those in $\tilde\lambda_i$ that are generated by Fourier transforming to momentum space to enforce
\begin{gather*}
\sum\limits_{i\in L}\lambda_i\tilde\lambda_i -\lambda_\bullet\sum\limits_{i\in L}t_i\tilde\lambda_i = s^2\rho\sum\limits_{i\in L}
t_i\tilde\lambda_i\frac{1}{u_{iL}} + \mathcal{O}(s^4),
\end{gather*}
where $\rho$ is the $\lambda$-component of a~map coef\/f\/icient that limits onto $\Sigma_R$.
This shows that, as in~\eqref{factorres},~\eqref{twistorres}, a~factorization channel in momentum space corresponds to a~nodal curve in twistor
space, with the same parameter $s^2$ governing both degenerations.

We can account for the various factors of~$u^{d_L}$ and~the rescalings in~\eqref{rescaledZ} as follows.
Firstly, unlike in the $\mathcal{N}=4$ Calabi--Yau case the $\mathcal{N}=8$ measure is not invariant under the resca\-ling~\eqref{rescaledZ} of the
map coef\/f\/icients, but rather behaves as
\begin{gather}
\prod\limits_{c=0}^d \mathrm{d}^{4|8}\mathcal{Z}'_c = s^{-2d_L(d_L+1)}s^{-2d_R(d_R+1)} \times \mathrm{d}^{4|8}\mathcal{Z}_\bullet
\prod\limits_{a=1}^{d_L}\mathrm{d}^{4|8}\mathcal{Z}_a\prod\limits_{b=1}^{d_R}\mathrm{d}^{4|8}\mathcal{Z}_b.
\label{Zmeasabsorbs}
\end{gather}
Secondly, bearing in mind that the wavefunctions and~matrix elements each depend homogeneously on $\mathcal{Z}(u)$, we can treat the map purely
as the terms in parentheses in~\eqref{LRmaps} provided we also make the replacements
\begin{gather}
h_i(\mathcal{Z}(u_i)) \to u_i^{2d_L} h_i(\mathcal{Z}(u_i)),
\qquad
\Phi_{ij} \to (u_iu_j)^{d_L} \Phi_{ij},
\qquad
\widetilde{\Phi}_{ij}\to \frac{1}{(u_iu_j)^{d_L}}\widetilde{\Phi}_{ij}.
\label{uabsorb}
\end{gather}
We do this henceforth.
In terms of the new coordinates, the product of the replaced wavefunctions becomes{\samepage
\begin{gather}
\prod\limits_{i=1}^n u_i^{2d_L}h_i(\mathcal{Z}(u_i)) = s^{2d_L(n_L-n_R)}\prod\limits_{i\in L}
\frac{h_i(\mathcal{Z}(u_{iL}))}{u_i^{2d_L}}\prod\limits_{j\in R} u_{jR}^{2d_L}h_j(\mathcal{Z}(u_{jR}))
\label{wavefunu}
\end{gather}
to leading order in $s^2$.}

The node itself is mapped to $\mathcal{Z}_\bullet\in\mathbb{PT}$.
It will be convenient to be able to treat the node separately on the two curve components.
For this, we introduce a~factor
\begin{gather}
1 = \int\mathrm{D}^{3|8}\mathcal{Y}_\bullet\mathrm{D}^{3|8}\mathcal{Z}\frac{\mathrm{d} t}{t}\frac{\mathrm{d}r}{r^5}
\bar\delta^{4|8}(\mathcal{Z} - t\mathcal{Z}_\bullet)\bar\delta^{4|8}(\mathcal{Z}-r\mathcal{Y}_\bullet)
\label{1asdeltas}
\end{gather}
into the integrand of~$\mathcal{M}_{n,d}$.
To understand this factor, f\/irst note that the powers of scaling parameters~$t$ and~$r$ in the measure are chosen to ensure the whole expression
has no weight in any of the three twistors.
The integrals can all be performed against the $\delta$-functions, which simply freeze $\mathcal{Y}_\bullet$ to $(t/r) \mathcal{Z}_\bullet$.
Now, whenever we describe a~particle in~$R$, we write the map as
\begin{gather*}
\mathcal{Z}(u_R) = \frac{t}{r}\left(\mathcal{Y}_\bullet +\sum\limits_{b=1}^{d_R}\mathcal{Y}_b u_R^b + \mathcal{O}\big(s^2\big)\right),
\end{gather*}
where
\begin{gather}
\mathcal{Y}_b:= \frac{r}{t}\mathcal{Z}_b
\label{YfromZb}
\end{gather}
are a~further rescaling of the $d_R$ map coef\/f\/icients $\mathcal{Z}_b$.
Note that we do not rescale the $d_L$ coef\/f\/icients $\mathcal{Z}_a$.
Pulling out this factor of~$r/t$ from all the wavefunctions $h_j(\mathcal{Z})$ with $j\in R$, and~from all the rows and~columns of~$\Phi$
and~$\widetilde{\Phi}$ corresponding to particles in~$R$, and~also changing the $\mathcal{Z}_b$ measure into that for $\mathcal{Y}_b$ leads to
a~factor of~$(r/t)^2$.
In the original formula~\eqref{N=8}, we divided by vol$(\mathbb{C }^*)$ to account for an overall rescaling of the map coef\/f\/icients.
However, as a~consequence of~\eqref{YfromZb}, the new $d_R$ map coef\/f\/ients $\mathcal{Y}_b$ are no longer locked to scale like
$\{\mathcal{Z}_\bullet,\mathcal{Z}_a\}$ but instead are locked to scale like $\mathcal{Y}_\bullet$.
This factor combines beautifully with the factors in the measure~\eqref{1asdeltas} to convert those integrals into our standard
$\bar\delta^{3|8}$'s of homogeneity $+2$ in each entry.
These $\delta$-functions can thus be treated as `external data' for the node.
Thus, as in~\eqref{twistorres}, as $s^2\to0$ the map degenerates into two independent maps from $(n_{L,R}+1)$-pointed curves $\Sigma_{L,R}$,
each described by $d_{L,R}+1$ twistor coef\/f\/icients, with a~point $\bullet_{L,R}$ on each curve mapped to the same point $\mathcal{Z}$ in the
target space.
The f\/inal integral $\mathrm{D}^{3|8}\mathcal{Z}$ allows this twistor to be anywhere, just as in the residue calculation~\eqref{twistorres}.

Now that we have described the degeneration, we must show that~\eqref{N=8} has a~simple pole there, with the correct residue.
Our f\/irst aim is to show that to leading order in $s^2$, the matrices~$\Phi$ and~$\widetilde{\Phi}$ become block diagonal so that their
determinants naturally factor into a~product of determinants for $\Phi_{L,R}$ and~for $\widetilde{\Phi}_{L,R}$.
Consider f\/irst~$\Phi$ and~assume that we choose the $\tilde d+1$ reference points~$p_r$ on the diagonal in~\eqref{phidef} so that $\tilde
d_{L,R}$ limit onto $\Sigma_{L,R}$, where
\begin{gather*}
\tilde d_{L,R}:= (n_{L,R} + 1) - d_{L,R} -2.
\end{gather*}
so that
\begin{gather*}
\tilde d= \tilde d_L+\tilde d_R.
\end{gather*}
The remaining marked point is chosen to be the node, viewed as being on the {\it right} when we consider diagonal elements $\Phi_{ii}$ with
$i\in L$, and~on the {\it left} for diagonal elements $\Phi_{jj}$ with $j\in R$.
Specif\/ically, we have{\samepage
\begin{gather*}
\Phi_{ij} = \frac{\langle \lambda(u_i)\lambda(u_j)\rangle}{(ij)}(u_iu_j)^{d_L},
\\
\Phi_{ii} = -\sum\limits_{j\neq i}\Phi_{ij} \prod\limits_{l=1}^{\tilde{d}_L}\frac{(p_l j)}{(p_l i)}
\prod\limits_{r=1}^{\tilde{d}_R}\frac{(p_r j)}{(p_r i)}\frac{(\bullet j)}{(\bullet i)}
\frac{\prod\limits_{k\neq i}(ki)}{\prod\limits_{m\neq j}(mj)}
\end{gather*}
in terms of the original coordinates, where we have accounted for the factors in~\eqref{uabsorb}.}

Using~\eqref{(ij)LR} to transform to the limit coordinates, we f\/ind that~$\Phi$ can be written as
\begin{gather}
{\Phi}_{ij} = \left.\frac{\langle \lambda(u_i)\lambda(u_j)\rangle}{(u_i-u_j)} \frac{s^{2d_L-1}}{(u_iu_j)^{d_L-1}}\right|_{L},
\nonumber
\\
{\Phi}_{ii} =-\sum\limits_{\substack{j\in L\cup\bullet\\j\neq i}}
\left({\Phi}_{ij}\left(\frac{u_j}{u_i}\right)^{d_L-1}
\prod\limits_{l=1}^{\tilde{d}_L}\frac{(u_{l}-u_{j})}{(u_{l}-u_{i})}\frac{(u_\bullet - u_{j})}{(u_\bullet-u_{i})}
\frac{\displaystyle{\prod\limits_{\substack{k\in L\cup\bullet\\k\neq i}}}(u_{k}-u_{i})}
{\displaystyle{\prod\limits_{\substack{m\in L\cup\bullet\\m\neq j}}}(u_{m}-u_{j})}\right)_{L}+\mathcal{O}\big(s^2\big)
\label{PhiL}
\end{gather}
when $i,j\in L$, and~where the subscript~$L$ on means we are using limiting coordinates appropriate for~$L$ throughout.
Similarly
\begin{gather}
{\Phi}_{ij} = \left.\frac{\langle \lambda(u_{i})\lambda(u_{j})\rangle}{u_{i}-u_{j}}\frac{(u_iu_j)^{d_L}}{s^{2d_L-1}}\right|_{R},
\nonumber
\\
{\Phi}_{ii} =-\sum\limits_{\substack{j\in R\cup\bullet
\\
j\neq i}} \left({\Phi}_{ij}\left(\frac{u_i}{u_j}\right)^{d_L}
\prod\limits_{r=1}^{\tilde{d}_R}\frac{(u_{r}-u_{j})}{(u_{r}-u_{i})}\frac{(u_\bullet-u_{j})}{(u_\bullet-u_{i})}
\frac{\displaystyle{\prod\limits_{\substack{k\in R\cup\bullet
\\
k\neq i}}}(u_{k}-u_{i})}{\displaystyle{\prod\limits_{\substack{m\in R\cup\bullet
\\
m\neq j}}}(u_{m}-u_{j})}\right)_{R}+\mathcal{O}\big(s^2\big)
\label{PhiR}
\end{gather}
when $i,j\in R$ and~again we use the~$R$ limiting coordinates.
Once we extract a~power of~$1/u_{iL}^{d_L-1}$ from each row and~column of~\eqref{PhiL}, a~power of~$u_{iR}^{d_L}$ from each row and~column
of~\eqref{PhiR} and~powers of~$s$ from both, these matrices are exactly of the form $\Phi_{L,R}$ for the subamplitudes.
Note that in both cases, we have extended the sum on the diagonal term to include the node (located at $u_\bullet=0$ in our coordinates).
This is possible because the choice of the node as a~reference point means this term is zero.
While $\Phi_{L,R}$ as given here are $n_{L,R}\times n_{L,R}$ matrices (rather than $(n_{L,R}+1)\times(n_{L,R}+1)$ matrices), they still have the
expected rank $d_{L,R}$, because in each case we were forced to choose one of the reference points to be the node.
It is as if the row and~column corresponding to the internal particle have `already' been removed.

The of\/f-block-diagonal terms $\Phi_{ij}$ with $i\in L$, $j\in R$ are of the same order in $s^2$ as the~$R$ block diagonal ones in~\eqref{PhiR}.
Therefore, the leading term in the reduced determinant comes from the block diagonal terms.
After also changing variables $u\to u_{L,R}$ in the Vandermonde determinants, a~straightforward but somewhat tedious calculation shows that
\begin{gather}
|\Phi|' = \frac{s^{(2d_L-1)(d_L-d_R)}s^{d_R(d_R-1)}}{s^{-2d_Ld_R}s^{d_L(d_L-1)}} |\Phi_L|'|\Phi_R|' + \text{higher order},
\label{Phifactorize}
\end{gather}
exactly as required for a~product of two subamplitudes, times an overall power of~$s$.

In an exactly parallel computation, transforming $\widetilde{\Phi}$ into the $L$, $R$ coordinates shows that
\begin{gather}
|\widetilde{\Phi}|' = \frac{s^{d_R(d_R+1)}s^{2(d_R+1)(d_L+1)}}{s^{d_L(d_L+1)}s^{(2d_L+1)(\tilde d_L-\tilde d_R)}}
|\widetilde{\Phi}_L|'|\widetilde{\Phi}_R|'\frac{\prod\limits_{i\in L} u_{iL}^{2(d_L+1)}}{\prod\limits_{j\in R}u_{jR}^{2d_L}}
\label{tPhifactorize}
\end{gather}
to leading order in $s^2$.
Once again the matrices $\widetilde{\Phi}_L$ and~$\widetilde{\Phi}_R$ are precisely as they should be for the left and~right subamplitudes,
where again we choose the node as one of the reference points.

After these somewhat lengthy calculations, we are f\/inally in position to compute the residue of~$\mathcal{M}_{n,d}$ on the boundary of the
moduli space corresponding to a~factorization channel.
First, collecting powers of~$s^2$ from equations~\eqref{measurewiths},~\eqref{Zmeasabsorbs},~\eqref{wavefunu},~\eqref{Phifactorize}
and~\eqref{tPhifactorize}, a~near miraculous cancellation occurs, leaving simply
\begin{gather*}
\mathrm{d} s^2\left(\frac{1}{s^2}(\cdots) + \mathcal{O}\big(s^0\big)\right),
\end{gather*}
showing that the integrand of~\eqref{N=8} indeed has a~simple pole on boundary divisors in the moduli space.
Combining all the pieces, the residue of this simple pole is
\begin{gather*}
\int\mathrm{D}^{3|8}\mathcal{Z}\left[ \int\mathrm{d}\mu_L\frac{\mathrm{d}^{4|8}\mathcal{Z}_\bullet
\prod\limits_{a=1}^{d_L}\mathrm{d}^{4|8}\mathcal{Z}_a}{{\rm vol}(\mathbb{C }^*)}|\Phi_L|'|\widetilde{\Phi}_L|'\prod\limits_{i\in L}
h_i(\mathcal{Z}(u_{iL})) \bar\delta^{3|8}(\mathcal{Z}(u_{\bullet L}),\mathcal{Z}) \right.
\\
\qquad
\times\left.\int\mathrm{d}\mu_R\frac{\mathrm{d}^{4|8}\mathcal{Y}_\bullet
\prod\limits_{b=1}^{d_R}\mathrm{d}^{4|8}\mathcal{Y}_b}{{\rm vol}(\mathbb{C }^*)}|\Phi_R|'|\widetilde{\Phi}_R|'
\prod\limits_{j\in R} h_j(\mathcal{Z}(u_{jR}))\bar\delta^{3|8}(\mathcal{Z}(u_{\bullet R}),\mathcal{Z}) \right],
\end{gather*}
or~in other words exactly the residue
\begin{gather*}
\int\mathrm{D}^{3|8}\mathcal{Z}\mathcal{M}_L(\{\mathcal{Z}_{i\in L}\},\mathcal{Z})\mathcal{M}_R(Z,\{\mathcal{Z}_{j\in R}\})
\end{gather*}
of the gravitational scattering amplitude.

We have now shown that $\mathcal{M}_{n,d}$ as given by equation~\eqref{N=8} produces the correct seed amplitudes for BCFW recursion, has the
correct $1/z^2$ decay as the BCFW shift parameter $z\to\infty$ and~has a~simple pole on any physical factorization channel, with residue
correctly given by the product of two subamplitudes on either side of the factorization, integrated over the phase space of the intermediate
state.

The only remaining thing to check is that in momentum space, $\mathcal{M}_{n,d}$ has no unwanted {\it unphysical} poles.
This is straightforward.
A simple dimension count of integrals versus constraints shows that, as for Yang--Mills~\cite{Roiban:2004yf}, when evaluated on momentum
eigenstates, $\mathcal{M}_{n,d}$ is inevitably a~rational function of the spinor momenta.
Thus the only possible singularities are poles.
Any unphysical poles in $\mathcal{M}_{n,d}$ which carry some helicity weight would be detected by taking one of the external momenta to become
soft.
Unphysical ``multiparticle" poles, i.e.\ poles that carry no helicity weight, would also be detected by sequentially taking many particles
to become soft.
However, the soft limits of~$\mathcal{M}_{n,d}$ have recently been checked to agree with those of gravity~\cite{Bullimore:2012parity}.
We therefore conclude that $\mathcal{M}_{n,d}$ indeed obeys the correct BCFW recursion relation, and~have thus demonstrated that it computes all
tree amplitudes in $\mathcal{N}=8$ supergravity.

\section{Parity invariance}\label{sec:parity}

One of the pleasing features of using~\eqref{N=8} to describe gravitational scattering amplitudes is that the way these amplitudes break
conformal symmetry becomes completely explicit: it arises purely from the inf\/inity twistors $\langle\;,\,\rangle$ and~$[\;,\,]$ and~in~$\Phi$
and~$\widetilde{\Phi}$, respectively.
On the other hand, parity transformations are not manifest, because parity exchanges twistor space with the dual twistor space.
For example, the twistor space $\mathbb{CP}^{3|\mathcal{N}}$ of conformally f\/lat space-time is exchanged with the dual projective space
${\mathbb{CP}^{3|\mathcal{N}}}^*$.
On the original $\mathcal{Z}$ twistor space, $[\;,\,]$ is a~dif\/ferential operator while $\langle\;,\,\rangle$ is multiplicative, so the role
of these brackets are interchanged under parity.
We see this change of roles quite transparently at the level of amplitudes: a~parity transformation f\/lips the helicities of all external states,
so it exchanges $d\leftrightarrow\tilde d$, and~one of the key observations of \cite{Cachazo:2012kg} was that the~$n$-particle N$^{d-1}$MHV
amplitude $\mathcal{M}_{n,d}$ is a~monomial of degree~$d$ in $\langle\;,\,\rangle$ and~of degree $\tilde d$ in $[\;,\,]$.
This strongly suggests that the determinants of~$\Phi$ and~$\widetilde{\Phi}$, which hitherto have seemed very dif\/ferent, are naturally parity
conjugates of each other.
Let's now see this explicitly.

Acting on either momentum or~twistor eigenstates, the matrix~$\Phi$ has elements
\begin{gather*}
\Phi_{ij} = \frac{\langle ij\rangle}{(ij)} \frac{1}{t_it_j},
\qquad
\Phi_{ii} =-\sum\limits_{j\neq i}\Phi_{ij}\prod\limits_{r=0}^{\tilde d}\frac{(jp_r)}{(ip_r)}\frac{\prod\limits_{k\neq i}
(ik)}{\prod\limits_{l\neq j} (jl)}.
\end{gather*}
To bring this to the form of~$\widetilde{\Phi}$, consider making the change of variables $t_i \to s_i$, def\/ined by
\begin{gather}
t_i s_i:= \frac{1}{\prod\limits_{j\neq i}(ij)}.
\label{parityscale}
\end{gather}
This transformation of the scaling parameters played a~key role in studying the behaviour of the connected prescription for $\mathcal{N}=4$ SYM
under parity~\cite{Roiban:2004yf,Witten:2004cp}; its relation to a~parity transformation will be reviewed below.
Under this change of variables, we f\/ind
\begin{gather}
\Phi^{(d)}(\langle\;,\,\rangle, t ) = A \circ \widetilde{\Phi}^{({\tilde d})}(\langle\;,\, \rangle, s) \circ A,
\label{parityPhi}
\end{gather}
where $\Phi^{(d)}$ is our usual~$\Phi$ matrix on a~degree~$d$ curve\footnote{The degree af\/fects the matrices only through the diagonal
elements.}, and~$\widetilde{\Phi}^{(\tilde d)}(\langle\;,\,\rangle, s)$ is the $\widetilde{\Phi}$ matrix appropriate for a~degree $\tilde d$
curve.
We also make the replacement $[\;,\,]\to\langle\;,\,\rangle$ in $\widetilde{\Phi}^{(\tilde d)}$.
Finally, $A$~is the diagonal matrix whose $j^{\rm th}$ entry is the product $\prod\limits_{k\neq j}(jk)$.
Acting with~$A$ as in~\eqref{parityPhi} multiplies the rows and~columns of~$\widetilde{\Phi}$ by this product, which accounts for the
denominator in~\eqref{parityscale}.

In equation~\eqref{detprimemultiply}, we saw that ${\det}'(\Phi)$ and~${\det}'(\widetilde{\Phi})$ behave just as usual determinants under matrix
multiplication.
In the present case we have
\begin{gather}
{\det}'\left(\Phi^{(d)}(\langle\;,\,\rangle, t )\right) = {\det}'\left(A \circ \widetilde{\Phi}^{({\tilde d})}(\langle\;,\, \rangle, s) \circ
A\right) =(\det A)^2 {\det}'\left(\widetilde{\Phi}^{({\tilde d})}(\langle\;,\, \rangle, s)\right)
\nonumber
\\
\hphantom{{\det}'\left(\Phi^{(d)}(\langle\;,\,\rangle, t )\right)}{}
 =|1\dots n|^{4}{\det}'\left(\widetilde{\Phi}^{({\tilde d})}(\langle\;,\, \rangle, s)\right).
\label{detprimeparity1}
\end{gather}
Similarly, if we start from $\widetilde{\Phi}^{(d)}([\;,\,],t)$ and~make the same change of variables~\eqref{parityscale}, then reading this
equation backwards gives
\begin{gather*}
{\det}'\left(\widetilde{\Phi}^{(d)}([\;,\,], t )\right) = |1\dots n|^{-4}{\det}'\left(\Phi^{({\tilde d})}([\;,\,], s)\right)
\end{gather*}
so that the product is
\begin{gather*}
{\det}'\left(\Phi^{(d)}(\langle\;,\,\rangle, t )\right){\det}'\left(\widetilde{\Phi}^{(d)}([\;,\,], t)\right)
={\det}'\left(\widetilde{\Phi}^{({\tilde d})}(\langle\;,\, \rangle, s)\right){\det}'\left(\Phi^{({\tilde d})}([\;,\,], s)\right),
\end{gather*}
with no extra factors.
Note that the roles of~$\Phi$ and~$\widetilde{\Phi}$ have been exchanged, along with the exchanges $\langle\;,\,\rangle \leftrightarrow[\;,\,]$
and~$d \leftrightarrow\tilde d$.

As mentioned above, in~\cite{Roiban:2004yf,Witten:2004cp} it was shown that the parity transformation of all the other factors in the
$\mathcal{N}=4$ SYM tree amplitudes $\mathcal{A}_{n,d}$ conspire to produce the transformation~\eqref{parityscale} of scaling parameters.
In $\mathcal{N}=4$ SYM, the measure for the scaling parameters themselves behaves as
\begin{gather*}
\frac{\mathrm{d} t_i}{t_i} \propto \frac{\mathrm{d} s_i}{s_i}
\end{gather*}
under~\eqref{parityscale}, with a~proportionality factor that is cancelled by the transformation of the fermions.
In $\mathcal{N}=8$ supergravity, the scaling parameters' measure is
\begin{gather*}
\frac{\mathrm{d} t_i}{t_i^3}\propto \mathrm{d} s_i\, s_i
\end{gather*}
instead, but now the transformation of the $\mathcal{N}=8$ fermions provides an extra factor of~$1/s_i^4$.
Exactly the same arguments as given in~\cite{Roiban:2004yf,Witten:2004cp} thus establish the parity symmetry of our formulation
of~$\mathcal{M}_{n,d}$.
Rather than simply repeat those arguments verbatim, we instead make parity manifest by recasting the integral~\eqref{N=8} in terms of the link
variables introduced in~\cite{ArkaniHamed:2009si} for $\mathcal{N}=4$ SYM.

\section{Gravity and~the Grassmannian}\label{sec:link}

The aim of this section is to write the tree amplitude $\mathcal{M}_{n,d}$ as an integral over the Grassmannian G$(k,n)$ (with $k=d+1$) along
the lines of~\cite{ArkaniHamed:2009dg,Bourjaily:2010kw,Bullimore:2009cb,Dolan:2009wf,Dolan:2011za,Nandan:2009cc,Spradlin:2009qr} for the
connected prescription of~$\mathcal{N}=4$ SYM.
The most obvious reason to perform this transformation is that as an integral over G$(k,n)$, all $\delta$-function constraints involving
external data become linear in the variables and~hence trivial to perform.
The price for such a~simplif\/ication is that the number of integrations variables is larger than before.
The dif\/ference in the number of variables is $(k-2)(n-k-2)$ and~hence the amplitude becomes a~multidimensional contour integral over that many
variables.

In $\mathcal{N}=4$ super Yang--Mills something remarkable happens: repeated applications of the global residue theorem transform the integral
into one where all variables can be solved for from linear systems
of equations~\cite{ArkaniHamed:2009dg,Bourjaily:2010kw,Dolan:2009wf,Dolan:2010xv,Dolan:2011za,Nandan:2009cc, Spradlin:2009qr}.
Computationally this is a~major advantage, but it also gives a~conceptual advantage because individual residues computed after the application
of the global residue theory coincide precisely with BCFW terms and~hence, in Yang--Mills, leading singularities of the theory.
One can then write down a~generating function for all leading singularities~\cite{ArkaniHamed:2009dn,Mason:2009qx} that control the behaviour
of the theory to all orders in perturbation theory and~which has allowed the development of recursion relations for the all loop
integrand~\cite{Arkani-Hamed:2010kv}.

These remarkable properties of the Grassmannian formulation of~$\mathcal{N}=4$ SYM should provide suf\/f\/icient motivation to explore the same
avenues in gravity.

It is important to realize that the existence of a~Grassmannian formulation {\it per se} has nothing to do with $\mathcal{N}=4$ SYM, or~Yangian
invariance, or~even twistors.
Rather, it is a~completely general consequence of dealing with degree~$d$ holomorphic maps from an~$n$-pointed rational curve.
To see this~\cite{Bullimore:2009cb}, recall that we can describe the map $\mathcal{Z}:\mathbb{CP}^1\to\mathbb{PT}$ by picking a~basis
$\{P_0(\sigma),\ldots,P_d(\sigma)\}$ of~$d+1$ linearly independent degree~$d$ polynomials in the worldsheet coordinates and~expanding
\begin{gather*}
Z(\sigma) = \sum\limits_{a=0}^d\mathcal{Z}_a P_a(\sigma).
\end{gather*}
The space of such polynomials is $H^0(\mathbb{CP}^1,\mathcal{O}(d))\cong\mathbb{C }^k$.
Given~$n$ marked points on the worldsheet, we would like to def\/ine a~natural embedding of this $\mathbb{C }^k$ into $\mathbb{C }^n$ by
`evaluating' each of the $P_a(\sigma)$ at each of the marked points.
This can be done once we f\/ix a~scale for $\sigma$ at each marked point~-- in other words, once we pick a~trivialization of~$\mathcal{O}(d)$ at
each of the $\sigma_i$.
This is exactly the role of the scaling parameters $t_i$.
Thus, for every choice of~$n$ marked points and~$n$ scaling parameters, our map def\/ines a~$k$-plane in $\mathbb{C }^n$, i.e.\
a~point in the Grassmannian G$(k,n)$.

As we integrate over the moduli space of rational maps, we sweep out a~$2(n-2)$-dimensional subvariety of~G$(k,n)$.
This dimension arises as
\begin{gather}
2(n-2) = (n-3) + (n)+ (- 1),
\label{parametercount}
\end{gather}
where $(n-3)$ parameters come from the locations of the marked points up to worldsheet SL$(2;\mathbb{C })$ transformations, a~further~$n$
parameters are the scaling parameters $t_i$ and~we lose one parameter from overall rescaling.
(Equivalently, we have $2n$ parameters from both components of the worldsheet coordinates $\sigma_i^{\underline\alpha}$, minus four from the
quotient by GL$(4;\mathbb{C })$.)

The precise subvariety we obtain may be characterized as follows~\cite{ArkaniHamed:2009dg, Bullimore:2009cb}.
The map from the worldsheet to the space of degree~$d$ polynomials, considered up to an overall scale, is of course the Veronese map
\begin{gather}
\mathcal{V}: \ \mathbb{CP}^1\to\mathbb{CP}^d.
\label{Veronese}
\end{gather}
The subvariety of the Grassmannian we sweep out is therefore def\/ined by the condition that the~$n$ dif\/ferent~$k$-vectors we get by evaluating
our polynomials do not simply span a~$k$-plane through the origin in $\mathbb{C }^n$, or~equivalently a~$\mathbb{CP}^d\subset\mathbb{CP}^{n-1}$
but rather lie in the image of the Veronese map to that $\mathbb{CP}^d$.
As shown in~\cite{ArkaniHamed:2009dg,Bourjaily:2010kw,Dolan:2009wf,Dolan:2010xv,Dolan:2011za,Nandan:2009cc, Spradlin:2009qr} and~obtained again
below, this condition amounts to the vanishing of~$(d-1)(\tilde d-1)$ quartics in the Pl{\"u}cker coordinates of the Grassmannian.
Note that
\begin{gather*}
\dim{\rm G}(d+1,n) - (d-1)(\tilde d-1) = 2(n-2)
\end{gather*}
giving the dimension expected from~\eqref{parametercount}.
On transforming to momentum space, the external data specif\/ies $2(n-2)$ divisors in G$(k,n)$ def\/ined by those~$k$-planes in $\mathbb{C }^n$ that
contain the 2-plane specif\/ied by the $\lambda_i$ and~are orthogonal to the 2-plane specif\/ied by the $\tilde\lambda_i$.
The intersection number of these divisors with the Veronese subvariety
is believed to be $\left\langle\substack{n-3\\k-2}\right\rangle$, where $\langle {\substack{p\\q}}\rangle$
is the $(p,q)$ Eulerian number~\cite{Roiban:2004yf}.

All the above features of the Grassmannian formulation should thus be common to both $\mathcal{N}=4$ super Yang--Mills and~$\mathcal{N}=8$
supergravity, purely as a~consequence of their having a~description in terms of degree~$d$ rational maps to twistor space.
Of course, the detailed form of the measure on the Grassmannian will be dif\/ferent in the two cases, coming from the external wavefunctions,
and~from the Parke--Taylor worldsheet denominator in Yang--Mills and~from $|\Phi|'$ and~$|\widetilde{\Phi}|'$ in gravity.

Let us now construct the Grassmannian formulation of tree amplitudes in $\mathcal{N}=8$ supergravity.
We will choose our external wavefunctions to be either twistor or~dual twistor eigenstates.
More precisely we choose exactly $d+1$ of the wavefunctions to be
\begin{gather}
h_a(\mathcal{Z}(\sigma_a)) = \int\frac{\mathrm{d} t_a}{t_a^3}\bar\delta^{4|8}(\mathcal{Z}_a -t_a\mathcal{Z}(\sigma_a))
\label{twistoreig}
\end{gather}
that have support only when $\sigma_a\in\Sigma$ is mapped to $\mathcal{Z}_a\in\mathbb{PT}$.
The remaining $\tilde d+1$ wavefunctions are chosen to be
\begin{gather*}
h_r(\mathcal{Z}(\sigma_r)) = \int \frac{\mathrm{d} t_r}{t_r^3} \exp\left(i t_r\mathcal{W}_r\cdot\mathcal{Z}(\sigma_r)\right)
\end{gather*}
that have plane-wave dependence on a~f\/ixed dual twistor $\mathcal{W}_r$.
We sometimes write components $\mathcal{W}_I=(\tilde\mu^\alpha,\tilde\lambda_{\dot\alpha},\psi_A)$ dual to the components
$\mathcal{Z}^I=(\lambda_\alpha,\mu^{\dot\alpha},\chi^A)$ of the original twistors.
Notice that both types of wavefunction have homogeneity $+2$ in $\mathcal{Z}(\sigma)$, as required for an $\mathcal{N}=8$ multiplet on twistor
space.
To recover the momentum space amplitude from these twistorial amplitudes, one Fourier transforms\footnote{The Fourier transform applies most
directly in (2,2) space-time signature, and~should more properly be understood as a~contour integral in other signatures.}
$\mu_a\to\tilde\lambda_a$ in twistor variables $\mathcal{Z}_a$ and~$\tilde\mu_r\to\lambda_r$ in the $\mathcal{W}_r$ dual twistor variables.
Since $\mu$ and~$\tilde\mu$ appear only in the exponentials, this Fourier transform is straightforward.

The main virtue of these external wavefunctions is that they provide exactly enough $\delta$-functions to perform all the integrals over the map
$\mathcal{Z}$.
If we pick our basis of polynomials to~be
\begin{gather}
\left\{\prod\limits_{b\neq a}\frac{(\sigma b)}{(ab)}\right\}
\qquad
\text{for}
\quad
|a|=d+1,
\label{polybasis}
\end{gather}
we can describe the map by
\begin{gather}
\mathcal{Z}(\sigma) = \sum\limits_{a=1}^{d+1}\mathcal{Y}_a \prod\limits_{b\neq a}\frac{(\sigma b)}{(ab)}.
\label{mapzeros}
\end{gather}
Then for $a=1,\ldots, d+1$, we have simply $\mathcal{Z}(\sigma_a)=\mathcal{Y}_a$, so the $k\times n$ matrix is f\/ixed to be the identity matrix
in the~$k$ columns corresponding to the~$a$-type particles.
In other words, with this choice of basis, our~$k$-plane inside $\mathbb{C }^n$ will be represented by the matrix
\begin{gather}
C_{ai} =
\begin{cases}
c_{ar}
&
\text{when}
\quad
i=r,
\\
\delta_{ab}
&
\text{when}
\quad
i=b
\end{cases}
\label{Cmatrix}
\end{gather}
for some parameters $c_{ra}$.
These parameters are known as `link variables'~\cite{ArkaniHamed:2009si}.
Using a~dif\/ferent choice of basis for $H^0(\mathbb{CP}^1,\mathcal{O}(d))$ would lead to a~GL$(k;\mathbb{C })$ transformation of~$C$, but the
point it def\/ines in the Grassmannian remains invariant.
Note that, with the parametrization given in~\eqref{Cmatrix}, since $c_{ab}=\delta_{ab}$, the link variables can be thought of directly as
minors of~$C_{ai}$.

There is a~small subtlety in using~\eqref{mapzeros} to describe the map, because~\eqref{polybasis} is not quite a~basis for
$H^0(\mathbb{CP}^1,\mathcal{O}(d))$ since $P_a(\sigma)$ in~\eqref{polybasis} has weight $-d$ in $\sigma_a$.
We can absorb this by declaring that for each~$a$, $\mathcal{Y}_a$ likewise has weight $+d$ under a~rescaling of~$\sigma_a$, so that
$\mathcal{Y}_a$ really takes values in $\mathcal{O}_a(d)$.
With the Calabi--Yau $\mathcal{N}=4$ supertwistor space, this may be done without comment, but with $\mathcal{N}=8$ supersymmetry we acquire
a~Jacobian in the measure for the integration over the map, which becomes
\begin{gather}
\prod\limits_{a=0}^d\mathrm{d}^{4|8}\mathcal{Z}_a = \prod\limits_{a=0}^d\left[\mathrm{d}^{4|8}\mathcal{Y}_a\prod\limits_{b\neq a}(ab)^2\right]
\label{measurerescale}
\end{gather}
in terms of the map coef\/f\/icients in~\eqref{mapzeros}.
This Jacobian cancels\footnote{The Jacobian is in the numerator because of the four extra fermionic components.
One can check that it has homogeneity $+4$ in each $\sigma_a$.} the scaling of~$\mathrm{d}^{4|8}\mathcal{Y}_a$.
With this subtlety accounted for, the wavefunctions~\eqref{twistoreig} enforce $\mathcal{Y}_a = \mathcal{Z}_a/t_a$ allowing us to integrate out
the map directly.

We are left with a~contribution
\begin{gather}
\prod\limits_a\mathrm{d} t_a t_a
\prod\limits_r\frac{\mathrm{d} t_r}{t_r^3}
\exp\left(\sum\limits_{r,a}\mathcal{W}_r\cdot\mathcal{Z}_a\frac{t_r}{t_a}\prod\limits_{b\neq a}\frac{(rb)}{(ab)}\right)
\label{wvfnint}
\end{gather}
from the external wavefunctions, where the measure for the $t_a$'s includes a~factor of~$t_a^4$ from solving the $\delta$-functions for the
$\mathcal{Y}_a$'s.
The factors
\begin{gather*}
\frac{t_r}{t_a}\prod\limits_{b\neq a}\frac{(rb)}{(ab)} = \frac{t_r}{t_a} P_a(\sigma_r)
\end{gather*}
in the exponential are precisely the Grassmannian coordinates $c_{ra}$ that we obtain by the procedure described above.
The ratio $t_r/t_a$ in front of~$P_a(\sigma_r)$ def\/ines a~trivialization of~$\mathcal{O}(d)$ at $\sigma_r$, and~so sets a~meaningful scale for
this ratio of homogeneous coordinates.

The next step is to manipulate our main formula~\eqref{N=8} so as to write it purely in terms of the external data
$\{\mathcal{W}_r,\mathcal{Z}_a\}$ and~the Grassmannian minors $c_{ra}$ (i.e.\ `link variables'), treated as independent variables.
Many of the required steps follow in close parallel to the computations
of~\cite{ArkaniHamed:2009dg,Bourjaily:2010kw,Dolan:2009wf,Dolan:2010xv,Dolan:2011za,Nandan:2009cc, Spradlin:2009qr}
in $\mathcal{N}=4$ SYM.
Since we did not f\/ind these manipulations to be particularly enlightening, we have postponed them to Appendix~\ref{sec:transformtolink}.

The f\/inal result is that all tree amplitudes in $\mathcal{N}=8$ supergravity can be written as the Grassmannian integral
\begin{gather}
\mathcal{M}_{n,d}(\{\mathcal{W}_r,\mathcal{Z}_a\})
\nonumber
\\
\qquad{}
=\int \left[\prod\limits_{r,a} \frac{\mathrm{d} c_{ra}}{c_{ra}}\right]
\frac{D_{12}^{n-1n}}{\prod\limits_a c_{1a}c_{2a}\prod\limits_r c_{rn-1}c_{rn}}
\left[\prod\limits_{\substack{r\neq1,2\\a\neq n-1,n}} D_{12}^{n-1n}c_{1a}c_{2a}c_{rn-1}c_{rn}
\bar\delta\left(\mathcal{V}_{12r}^{an-1n}\right)\right]
\nonumber
\\
\qquad\quad{}
\times
\phi^{(d)}\left(\frac{\langle ab\rangle}{H_{12}^{ab}}\right)
\phi^{(\tilde d)}\left(\frac{[rs]}{H_{rs}^{n-1n}}\right)
\exp\left(\sum\limits_{r,a}c_{ra} \mathcal{W}_r\cdot\mathcal{Z}_a\right).
\label{N=8Gkn}
\end{gather}
Here,
$D$ and~$H$ are the quadratic polynomials
\begin{gather}
D_{rs}^{ab}:=\left|
\begin{matrix}
c_{ra}&c_{rb}
\\
c_{sa}&c_{sb}
\end{matrix}
\right|
\qquad
\text{and}
\qquad
H_{rs}^{ab}:= \frac{D_{rs}^{ab}}{c_{ra}c_{rb}c_{sa}c_{sb}}
\label{DHdef}
\end{gather}
in the minors $c_{ra}$, while the $\bar\delta$-functions in the sextic polynomials
\begin{gather*}
\mathcal{V}_{12r}^{an-1n}:= \left|
\begin{matrix}
c_{1a}c_{1n-1} & c_{1n-1}c_{1n} & c_{1n}c_{1a}
\\
c_{2a}c_{2n-1} & c_{2n-1}c_{2n} & c_{2n}c_{2a}
\\
c_{ra}c_{rn-1} & c_{rn-1}c_{rn} & c_{rn}c_{ra}
\end{matrix}
\right|
\end{gather*}
restrict the support of the integral to the subvariety of~G$(k,n)$ def\/ined by the Veronese map~\eqref{Veronese}, as expected.
The functions $\phi^{(d)}$ and~$\phi^{(\tilde d)}$ represent the determinants ${\det}'(\Phi){\det}'(\widetilde{\Phi})$ from~\eqref{N=8}.
$\phi^{(d)}$ is def\/ined to be the determinant of the $d\times d$ symmetric matrix with elements
\begin{gather}
\phi_{ab} = \frac{\langle ab\rangle}{H_{12}^{ab}}
\qquad
\text{for}
\quad
a\neq b
\quad
\text{and}
\quad
a,b\neq n,
\qquad
\phi_{aa} =-\sum\limits_{b\neq a}\frac{\langle ab\rangle}{H_{12}^{ab}}
\label{dphidef}
\end{gather}
running over all of the $\mathcal{Z}_a$-type particles, except for one which without loss we take to be $\mathcal{Z}_n$.
(Note however that $\mathcal{Z}_n$ does appear in the diagonal entries.) Similarly, $\phi^{(\tilde d)}$ is the determinant of the $\tilde
d\times \tilde d$ symmetric matrix with elements
\begin{gather}
\phi_{rs} = \frac{[rs]}{H_{rs}^{n-1n}}
\qquad
\text{for}
\quad
r\neq s
\quad
\text{and}
\quad
r,s\neq 1,
\qquad
\phi_{rr} =-\sum\limits_{s\neq r}\frac{\langle rs\rangle}{H_{12}^{ac}}
\label{tildedphidef}
\end{gather}
running over all of the $\mathcal{W}_r$-type particles except for one which we take to be $\mathcal{W}_1$ (again ap\-pearing in the diagonal).
Notice that the inf\/inity twistor in the form $[\;,\,]$ sees only the $\mathcal{W}_r$'s (which contain $\tilde\lambda_r$'s), while the inf\/inity
twistor $\langle\;,\,\rangle$ sees only the $\mathcal{Z}_a$'s (containing $\lambda_a$'s).
These are therefore multiplicative operators in both cases.
Also notice that, as usual in the link representation, parity invariance is now completely manifest.

{\bf Note.}
As this manuscript was being prepared to be submitted~\cite{Bullimore:2012parity} and~\cite{He:2012link} appeared.
The former has overlap with the parity invariance proof given in Section~\ref{sec:parity}
while the latter overlaps with the link representation formula presented in Section~\ref{sec:link}.

\appendix

\section{Some properties of determinants}\label{sec:determinants}

In this appendix we give a~careful def\/inition of the determinants ${\det}'(\Phi)$ and~${\det}'(\widetilde{\Phi})$ that appear in~\eqref{N=8},
and~to gather a~few general results about such determinants.
All the material in here is standard mathematics.

The symmetric matrix~$\Phi$ def\/ines an inner product on an~$n$-dimensional vector space that we call~$V$.
The~$m$-dimensional kernel of~$\Phi$ is characterized by an $m\times n$ matrix of relations~$R$.
This is summarized in the sequence
\begin{gather*}
0\longrightarrow W \stackrel{R}\longrightarrow V \stackrel{\Phi}\longrightarrow V^*\stackrel {R^{\rm T}} \longrightarrow W^*\longrightarrow 0,
\end{gather*}
where $\Phi \circ R=0$ and~$R^{\rm T}\circ\Phi=0$.
The sequence is exact if $\ker\Phi={\rm im} R$ and~$\ker R^{\rm T} = {\rm im}\Phi$ so that the matrices otherwise have maximal rank ($m$
for~$R$ and~$(n-m)$ for~$\Phi$).
If we are given top exterior forms $\epsilon$ and~$\varepsilon$ on~$V$ and~$W$ respectively, the determinant ${\det}'(\Phi)$ may be def\/ined in
an invariant way via the equation\footnote{Really, ${\det}'(\Phi)$ is the determinant of the whole sequence.}
\begin{gather}
\epsilon^{i_1 \cdots i_n}\epsilon^{j_1\cdots j_n}\Phi_{i_{m+1} j_{m+1}}\cdots \Phi_{i_{n}j_{n}} =: {\det}'(\Phi)\varepsilon^{r_1\cdots
r_m} R^{j_1}_{r_1}\cdots R^{j_m}_{r_m} \varepsilon^{s_1\cdots s_m} R_{s_1}^{i_1}\cdots R_{s_m}^{i_m}.
\label{detprimedef}
\end{gather}
That this identity is true for some ${\det}'(\Phi)$ follows from the fact that, while the left hand side is non-zero by the assumption on the
rank of~$\Phi$, it vanishes if contracted with any further copy of~$\Phi$.
Since the kernel is characterized by the~$m$ vectors~$R$, the~$m$ upstairs skew indices $i_1,\ldots,i_m$ and~$j_1,\ldots,j_m$ must each be
a~multiple of the $m^{\rm th}$ exterior power of the~$R$'s.
Choosing any values for these free indices, we immediately obtain the standard formula
\begin{gather}
{\det}'(\Phi) =\frac {|\Phi|^{i_1\dots i_m}_{j_1\cdots j_m}}{ \varepsilon^{r_1\cdots r_m} R^{j_1}_{r_1}\cdots R^{j_m}_{r_m}
\varepsilon^{s_1\cdots s_k} R_{s_1}^{i_1}\cdots R_{s_k}^{i_k}}.
\label{detprimestandard}
\end{gather}
The above argument shows this expression is independent of the choice of indices.

It will be useful to understand the behaviour of reduced determinants when~$\Phi$ is multiplied by a~non-singular $n\times n$ matrix~$A$.
If~$\Phi$ has kernel~$R$, then $\Phi A$ has kernel $A^{-1}R$.
We now replace~$\Phi$ by $\Phi A$ in the def\/inition~\eqref{detprimedef} and~multiplying through by~$m$ further~$A$'s.
On the left we obtain a~factor of the determinant of~$A$, while on the right the multiplication cancels the factors of~$A^{-1}$ in the~$m$-fold
exterior product of~$(A^{-1}R)$.
We therefore obtain simply
\begin{gather}
{\det}'(\Phi A) ={\det(A)} {\det}'(\Phi).
\label{detprimemultiply}
\end{gather}
In particular, provided~$A$ is non-singular, conjugation $\Phi\to A^{-1}\Phi A$ does not change the reduced determinant.

In our case, neither the map~$\Phi$ nor the vector spaces~$V$ and~$W$ are really f\/ixed, but depend on parameters such as the map to twistor
space and~the locations of the vertex operators.
Because we can rescale these parameters, there are no preferred top exterior forms $\epsilon$ or~$\varepsilon$.
The determinant ${\det}'(\Phi)$ is not really a~number, but a~section of the determinant line bundle $(\wedge ^nV^*)^2\otimes(\wedge^m W)^2$
over the space of parameters.
We need to check that the determinant line bundles def\/ined by~$\Phi$ and~$\widetilde{\Phi}$ combine with the rest of the factors in the
integrand to form a~canonically trivial bundle, so that the whole expression is invariant under local rescalings of the worldsheet and~map
homogeneous coordinates.

We can keep track of the behaviour under rescalings of the homogeneous coordinates $\sigma$ and~$\mathcal{Z}$ by def\/ining a~quantity to take
values in $\mathcal{O}_i(1)$ if it has homogeneity 1 under rescaling of~$\sigma_i$, and~values in $\mathcal{O}[1]$ if it has homogeneity 1 under
rescaling of~$\mathcal{Z}$.
Thus for example the relation $\lambda_i = t_i\lambda(\sigma_i)$ means that\footnote{No other factor in~\eqref{N=8} has non-trivial behaviour
under scalings of the external data, so for the purposes of this discussion, we can keep the external data f\/ixed.}
$t_i\in\mathcal{O}_i(-d)[-1]$.
The weights of~$\Phi_{ij}$ then identify~$\Phi$ as a~symmetric form on
\begin{gather}
V=\bigoplus_{i=1}^n \mathcal{O}_i(1-d)[-1]
\label{Videntify}
\end{gather}
so that~$\Phi$ gives a~pure number, invariant under rescalings, when evaluated on two elements of~$V$.
With the kernel of~$\Phi$ def\/ined by~\eqref{kerphi}, the map~$R$ is explicitly
\begin{gather}
R^j_{r}=\frac{\sigma_j^{\underline\alpha_1}\cdots\sigma_j^{\underline\alpha_{\tilde d+1}}}{\prod\limits_{i\neq j}(ij)}.
\label{Rexplicit}
\end{gather}
Since~$V$ is given by~\eqref{Videntify}, this identif\/ies~$W$ as
\begin{gather*}
W=\mathbb{C }^{n-d}\otimes\mathcal{O}[-1]\bigotimes_{i=1}^n\mathcal{O}_i(1),
\end{gather*}
where $\mathbb{C }^{n-d}$ is (dual to) the space $H^0(\Sigma,\mathcal{O}({\tilde d}+1))$ of degree ${\tilde d}+1$ polynomials in $\sigma$.
The determinant line bundle associated to~$\Phi$ is thus
\begin{gather*}
\big({\wedge}^nV^*\big)^2\otimes\big({\wedge}^m W\big)^2\cong \mathcal{O}[2d]\bigotimes_i\mathcal{O}_i(2n-2)
\end{gather*}
so that ${\det}'(\Phi)$ has homogeneity $2d$ in $\mathcal{Z}$ and~$2n-2$ in each of the~$n$ points $\sigma_i$.

Considering the exact sequence
\begin{gather*}
0\longrightarrow {\tilde W} \stackrel{{\tilde R}}\longrightarrow {\tilde V} \stackrel{\widetilde{\Phi}}\longrightarrow {\tilde V}^*\stackrel
{{\tilde R}^{\rm T}} \longrightarrow {\tilde W}^*\longrightarrow 0
\end{gather*}
for $\widetilde{\Phi}$, we can likewise identify
\begin{gather*}
\tilde V = \bigoplus_{i=1}^n\mathcal{O}_i(d+1)[1]
\qquad
\text{and}
\qquad
\tilde W = \mathbb{C }^{d+2}\otimes\mathcal{O}[1],
\end{gather*}
so that ${\det'}(\widetilde{\Phi})$ is a~section of the determinant line bundle
\begin{gather*}
\big({\wedge}^n{\tilde V}^*\big)^2\otimes\big({\wedge}^{d+2} {\tilde W}\big)^2\cong\mathcal{O}\big[{-}2{\tilde d}\,\big] \bigotimes_{i=1}^n\mathcal{O}_i(-2d-2).
\end{gather*}
Combining the two determinants and~the explicit Vandermonde factor in~\eqref{N=8} shows that
\begin{gather*}
\frac{{\det}'(\Phi){\det}'(\widetilde{\Phi})}{|1\dots n|^2} \in \mathcal{O}[4(d+1) - 2n]\bigotimes_{i=1}^n\mathcal{O}_i(-2d-2),
\end{gather*}
which correctly conspires to cancel the scaling $-4(d+1)$ of the measure $\prod\limits_{\rm a}\mathrm{d}^{4|8}\mathcal{Z}$ and~the wavefunctions
$\prod\limits_{i=1}^n(\sigma_i\mathrm{d}\sigma_i)h_i(\mathcal{Z}(\sigma_i))$.
The integrand of~\eqref{N=8} is thus projectively well-def\/ined.

Finally, let us comment on relation of the def\/inition of~${\det}'(\Phi)$ given here to that given in~\cite{Cachazo:2012kg}.
Here, the denominator in~\eqref{detprimestandard} involves
\begin{gather}
R^{i_1} \wedge R^{i_2}\wedge\dots\wedge R^{i_m} = \frac{| i_1\dots i_m |}{\prod\limits_{r=1}^m \left(\prod\limits_{k\neq i_r} (i_r k)\right)},
\label{Rdenom}
\end{gather}
where numerator of this expression is the Vandermonde determinant of all the worldsheet coordinates associated with the components of~$\Phi$
that are absent in~\eqref{detprimedef}, while the denominator comes from the factors in the denominator of~$R$ in~\eqref{Rexplicit}.
A little experimentation shows that when $d>1$~\eqref{Rdenom} can also be written as
\begin{gather*}
R^{i_1} \wedge R^{i_2}\wedge\dots\wedge R^{i_m} = \frac{|i_{m+1}\dots i_n|}{|1\dots n|},
\end{gather*}
where $|i_{m+1}\dots i_n|$ is the Vandermonde determinant corresponding to the components of~$\Phi$ that are {\it present}
in~\eqref{detprimedef}.
(For $d=0,1$ the exterior product of all the~$R$'s gives exactly $|1\dots n|^{-1}$.)
An exactly analogous statement is true for the cokernel
def\/ined by $R^{\rm T}$.
It is amusing to notice that the explicit factor of~$|1\dots n|^2$ in~\eqref{N=8} can thus be absorbed into ${\det}'(\Phi)$ if we def\/ine this as
\begin{gather*}
{\det}'(\Phi) = \frac {|\Phi|^{i_1\dots i_m}_{j_1\dots j_m}}{|i_{m+1}\dots i_n|\, |j_{m+1}\dots j_n|}
\end{gather*}
instead of by~\eqref{detprimestandard}.
This is the def\/inition that was used in~\cite{Cachazo:2012kg} and~it is often more convenient for explicit calculations.

\section{Transformation to the link variables}\label{sec:transformtolink}

In this appendix we explain how $\mathcal{M}_{n,d}(\{\mathcal{W}_r,\mathcal{Z}_a\})$ can be manipulated so as to be written as an integral over
the Grassmannian, gauged f\/ixed to the link representation.

With the aim of simplifying the argument of the exponentials in~\eqref{wvfnint} it is useful to replace the scaling parameters $(t_a,t_r)$ by
parameters $(S_a,T_r)$ def\/ined as in~\cite{Dolan:2009wf, Spradlin:2009qr} by
\begin{gather}
S_a:= \frac{1}{t_a \prod\limits_{b\neq a}(ab)}
\qquad
\text{and}
\qquad
T_r:= t_r\prod\limits_b (rb).
\label{STdef}
\end{gather}
Notice that
\begin{gather}
S_a = s_a\prod\limits_r(ar),
\label{Ss}
\end{gather}
where $s_a$ was given in~\eqref{parityscale}.
In terms of these $(S_a,T_r)$ variables,~\eqref{wvfnint} becomes
\begin{gather*}
\prod\limits_a \left[\frac{\mathrm{d} S_a}{S_a^3} \prod\limits_{b\neq a}\frac{1}{(ab)^2}\right]
\prod\limits_r\left[\frac{\mathrm{d}T_r}{T_r^3}\prod\limits_b(rb)^2\right]
\exp\left(\sum\limits_{r,a}\mathcal{W}_r\cdot\mathcal{Z}_a\frac{T_rS_a}{(ra)}\right),
\end{gather*}
where the $1/(ra)$ factor appears in the exponential because $(ra)$ is absent in~\eqref{wvfnint} but present in the def\/inition of~$T_r$.
The factor of~$\prod\limits_a\left[\prod\limits_{b\neq a}(ab)^{-1}\right]$ precisely cancels the Jacobian factor in~\eqref{measurerescale}
associated with our choice of basis polynomials.

An advantage of choosing $d+1$ wavefunctions to be of the form~\eqref{twistoreig} is that we may choose to remove rows and~columns from~$\Phi$
and~$\widetilde{\Phi}$ in such a~way that ${\det}'(\Phi)$ depends only on the $\mathcal{Z}_a$ while ${\det}'(\widetilde{\Phi})$ depends only on
the $\mathcal{W}_r$.
Using the identity~\eqref{detprimeparity1} we have
\begin{gather*}
\frac{{\det}'(\Phi){\det}'(\widetilde{\Phi})}{|1\dots n|^2} = |1\dots n|^2{\det}'\left(\widetilde{\Phi}^{({\tilde d})}(\langle\;,\, \rangle,
s)\right){\det}'\left(\widetilde{\Phi}^{(d)}([\;,\,],t)\right).
\end{gather*}
We then choose to remove from $\widetilde{\Phi}^{(\tilde d)}(\langle\;,\,\rangle,s)$ all the $\tilde d+1$ rows and~columns corresponding to the
$\mathcal{W}_r$ particles, together with the row and~column of one of the~$\mathcal{Z}_a$ particles.
Without loss of generality, we take this to be `$n$'.
The of\/f-diagonal elements of~$\widetilde{\Phi}^{(\tilde d)}(\langle\;,\,\rangle,s)$ are now independent of the external $\mathcal{W}_r$'s.
To remove the $\mathcal{W}_r$'s from the diagonal elements, we additionally choose the $\tilde d+1$ reference points $p_r$ to be the worldsheet
insertion points $\sigma_r$ of the $\mathcal{W}_r$ wavefunctions.
Then the remaining elements of~$\widetilde{\Phi}^{(\tilde d)}(\langle\;,\,\rangle,s)$ are
\begin{gather}
\widetilde{\Phi}^{(\tilde d)}_{ab} = \frac{\langle ab\rangle}{(ab)} S_aS_b\prod\limits_{r}\frac{1}{(ar)(br)},
\qquad
\widetilde{\Phi}^{(\tilde d)}_{aa} = -\sum\limits_{b\neq a}\frac{\langle ab\rangle}{(ab)}S_aS_b\prod\limits_{r} \frac{1}{(ar)^2},
\label{tPhitilded}
\end{gather}
where the sum runs only over the $\mathcal{Z}_b$ particles, and~we have used~\eqref{Ss}.

Similarly, in $\widetilde{\Phi}^{(d)}([\;,\,],t)$ we choose to remove the $d+2$ rows and~columns corresponding to the $\mathcal{Z}_a$ and~to, say,
$\mathcal{W}_1$.
In addition we must choose the $d+1$ reference points $p_a=\sigma_a$ so as to ensure $\mathcal{Z}_a$ does not arise on the diagonal.
The remaining elements of~$\widetilde{\Phi}$ are then
\begin{gather}
\widetilde{\Phi}_{rs} =\frac{[rs]}{(rs)} T_rT_s\prod\limits_b\frac{1}{(rb)(sb)},
\qquad
\widetilde{\Phi}_{rr} =-\sum\limits_{s\neq r}\frac{[rs]}{(rs)}T_rT_s \prod\limits_{b}\frac{1}{(rb)^2},
\label{tPhid}
\end{gather}
where again the sum runs only over the $\mathcal{W}_s$-type particles and~where we have used~\eqref{STdef}.

The next step is to examine the reduced determinants.
We can remove a~factor of~$\prod\limits_{a\neq 1}\!\!\left[\!\prod\limits_{r}(ab)^{-2}\!\right]\!$
from the reduced determinant of~\eqref{tPhitilded}
and~a~factor of~$\prod\limits_{r\neq 1}\left[\prod\limits_b (rb)^{-2}\right]$ from the reduced determinant of~\eqref{tPhid}.
For later convenience, we also multiply every remaining element of~\eqref{tPhitilded} by $T_1T_2/ (12)$ and~every remaining element
of~\eqref{tPhid} by $S_{n-1}S_n/(n-1\,n)$.
Collecting all the factors, the determinants become
\begin{gather}
\frac{{\det}'(\Phi){\det}'(\widetilde{\Phi})}{|1\dots n|^2} = \phi^{(d)}\left(\frac{\langle ab\rangle S_aS_bT_1T_2}{(12)(ab)}\right)
\phi^{(\tilde d)}\left(\frac{[rs]T_rT_sS_{n-1}S_n}{(rs)(n-1\,n)}\right)
\frac{(12)^d(n-1\,n)^{\tilde d}}{(T_1T_2)^d(S_{n-1}S_n)^{\tilde d}}
\nonumber
\\
\hphantom{\frac{{\det}'(\Phi){\det}'(\widetilde{\Phi})}{|1\dots n|^2} =}{}
\times
\frac{|1\dots n|^2}{|nr_0\dots r_{\tilde d}|^2|1a_0\dots a_d|^2}\frac{\prod\limits_r(rn)^2\prod\limits_a(1a)^2}{\prod\limits_{a,r}(ar)^4},
\label{collected}
\end{gather}
where $\phi^{(d)}$ is the determinant of the $d \times d$ symmetric matrix def\/ined in~\eqref{dphidef} and~$\phi^{(\tilde d)}$ is the determinant
of the $\tilde d\times\tilde d$ symmetric matrix def\/ined in~\eqref{tildedphidef}.
The last line of~\eqref{collected} combines with the factor of~$\prod\limits_{a,r}(ar)^2$ left over from the change of the measure $t_r\to T_r$
and~then cancels completely.

We now introduce the link variables
\begin{gather}
c_{ra}:=\frac{t_r}{t_a}\prod\limits_{b\neq a}\frac{(rb)}{(ab)} = \frac{T_rS_a}{(ra)}
\label{linkcdef}
\end{gather}
so that the argument of the exponential in~\eqref{wvfnint} becomes simply $\sum\limits_{a,r}c_{ar}\mathcal{W}_r\cdot\mathcal{Z}_a$.
We can treat the $c_{ra}$'s as $(d+1)(\tilde d+1)$ independent variables if we enforce the conditions~\eqref{linkcdef} by introducing further
$\delta$-functions into $\mathcal{M}_{n,d}$ via
\begin{gather*}
1=\prod\limits_{r,a}\mathrm{d} c_{ra}\bar\delta\left(c_{ra}-\frac{T_rS_a}{(ra)}\right).
\end{gather*}
At this point, almost all of the formula for the amplitude can immediately be written in terms of the~$c$'s and~the external data.
We f\/ind
\begin{gather*}
\mathcal{M}_{n,d}(\{\mathcal{W}_r,\mathcal{Z}_a\})
\\
\qquad{}
=\int\frac{\prod\limits_{i=1}^n(\sigma_i\mathrm{d}\sigma_i)}{{\rm vol\,GL(2;\mathbb{C })}}\prod\limits_r\frac{\mathrm{d}
T_r}{T_r^3}\prod\limits_{a}\frac{\mathrm{d} S_a}{S_a^3}
\left[\prod\limits_{r,a}\mathrm{d} c_{ra}\bar\delta\left(c_{ra}-\frac{T_rS_a}{(ra)}\right)\right]
\frac{(n-1\,n)^{\tilde d-d}}{(S_{n-1}S_n)^{\tilde d-d}}\left(H_{12}^{n-1n}\right)^d
\\
\qquad\quad{}
\times\phi^{(d)}\left(\frac{\langle ab\rangle}{H_{12}^{ab}}\right)\phi^{(\tilde d)}\left(\frac{[rs]}{H_{rs}^{n-1n}}\right)
\exp\left(\sum\limits_{r,a}c_{ra}\mathcal{W}_r\cdot\mathcal{Z}_a\right),
\end{gather*}
where we have def\/ined
\begin{gather*}
H_{rs}^{ab}:=\frac{1}{c_{ra}c_{sb}}-\frac{1}{c_{rb}c_{sa}} = \frac{c_{rb}c_{sa}-c_{ra}c_{sb}}{c_{ra}c_{rb}c_{sa}c_{sb}}.
\end{gather*}
as in~\eqref{DHdef}.

To reach our f\/inal form of the Grassmannian representation of gravitational tree amplitudes, depending exclusively on the $c_{ra}$'s
and~external data, we must perform the $(\sigma,T,S)$ integrals.
This is a~straightforward, if rather lengthy exercise.
We choose to f\/ix the ${\rm SL}(2;\mathbb{C })$ freedom by freezing $\sigma_1$, $\sigma_{n-1}$ and~$\sigma_n$ to some arbitrary values at the usual
expense of a~Jacobian $(1\,n-1)(n-1\,n)(n1)$, and~f\/ix the scaling by freezing $S_n=1$ (for which the Jacobian is $S_n=1$).
The integrals are then performed using $2n$ of the $(d+1)(\tilde d+1)$ $\delta$-functions, and~lead directly to~\eqref{N=8Gkn} given in the main
text.
Note in particular that the Veronese constraints $\bar\delta(\mathcal{V}_{12r}^{an-1n})$ that remain in~\eqref{N=8Gkn} arise simply from
repeatedly substituting the support of one $\delta$-function into another.

\section{Conventions}\label{sec:conventions}

Let us list our conventions.
We take $\mathbb{PT}$ to be the $\mathcal{N}=8$ supertwistor space $\mathbb{CP}^{3|8}$ with a~line~$I$ removed.
We use calligraphic letters to denote supertwistors, lowercase and~uppercase Roman indices to denote their four bosonic and~$\mathcal{N}$
fermionic components, respectively.
We often decompose the bosonic components into two 2-component Weyl spinors with dotted and~undotted Greek indices.
Thus $\mathcal{Z} = (Z^{\rm a},\chi^A) = (\lambda_\alpha,\mu^{\dot\alpha}, \chi^A)$.
External states are labelled by lowercase Roman indices from the middle of the alphabet $i,j,\ldots \in\{1,\ldots,n\}$.
We use $\sigma^{\underline{\alpha}}$ with $\underline{\alpha}, \underline{\beta}, \ldots\in\{1,2\}$
to denote a~homogeneous coordinate on the $\mathbb{CP}^1$ worldsheet.
We often choose italic letters from the beginning of the alphabet to run over the space of degree~$d$ polynomials in the worldsheet coordinates,
so $a,b,\ldots \in\{0,\ldots, d\}$.
It is also useful to separately allow $r,s,\ldots \in \{0,\ldots, \tilde d\}$.
We use $[\;,\, ]$ to denote dotted spinor contractions,
$\langle\;,\, \rangle$ for undotted contractions, and~$(\;,\, )$ for contractions of the
homogeneous coordinates $\sigma$ on the worldsheet.
When af\/f\/ine coordinates are more convenient, we will choose them so that $\sigma=(1,u)$.
We shall denote the data of external spinor supermomenta by $\Lambda=(\lambda_\alpha,\tilde\lambda_{\dot\alpha},\eta_A)$.

\subsection*{Acknowledgments}

This work is supported by the Perimeter Institute for Theoretical Physics.
Research at the Perimeter Institute is supported by the Government of Canada through Industry Canada and~by the Province of Ontario through the
Ministry of Research and Innovation.
The work of~FC is supported in part by the NSERC of Canada and~MEDT of Ontario.
LM is supported by a~Leverhulme Fellowship.

\pdfbookmark[1]{References}{ref}
\LastPageEnding

\end{document}